\documentclass[11pt]{article}
\usepackage[utf8]{inputenc}
\usepackage[english]{babel}
\usepackage{amsmath}
\usepackage{setspace}
\usepackage{graphicx}
\usepackage{bbm}
\usepackage[margin=1in]{geometry}
\usepackage{float}
\usepackage{multirow}
\usepackage{graphicx}
\usepackage{chngcntr}
\usepackage{hhline}
\usepackage{comment}
\usepackage{bigstrut}
\usepackage{natbib}
\usepackage{hyperref}
\usepackage{xcolor}
\definecolor{darkblue}{RGB}{0, 0, 139} % Define a custom dark blue color
\hypersetup{colorlinks=true,linkcolor=darkblue,citecolor=darkblue,urlcolor=darkblue,hypertexnames=true}

\title{Underlying Core Inflation with Multiple Regimes\thanks{
The author thanks Mikael Khan, Luis Uzeda, Geoffrey R. Dunbar, Kerem Tuzcuoglu, Tara M. Sinclair, Hélène Desgagnés, Edouard Djeutem, Jean-Marie Dufour, Philippe Goulet Coulombe, as well as several Bank of Canada seminar participants and several conference participants for useful comments.
This work was supported by the Fonds de recherche sur la société et la 
culture Doctoral Research Scholarships (B2Z).}}
\date{\today}
\author{Gabriel Rodriguez-Rondon\thanks{Mailing address: Department of 
Economics, McGill University, 855 Sherbrooke St. West Montreal, QC H3A 
2T7. e-mail: gabriel.rodriguezrondon@mail.mcgill.ca. Wed page: \url{https://grodriguezrondon.com}.}}

\begin{document}

\maketitle

\begin{abstract}
This paper introduces a new approach for estimating core inflation indicators based on common factors across a broad range of price indices. Specifically, by utilizing procedures for detecting multiple regimes in high-dimensional factor models, we propose two types of core inflation indicators: one incorporating multiple structural breaks and another based on Markov switching. The structural breaks approach can eliminate revisions for past regimes, though it functions as an offline indicator, as real-time detection of breaks is not feasible with this method. On the other hand, the Markov switching approach can reduce revisions while being useful in real time, making it a simple and robust core inflation indicator suitable for real-time monitoring and as a short-term guide for monetary policy. Additionally, this approach allows us to estimate the probability of being in different inflationary regimes. To demonstrate the effectiveness of these indicators, we apply them to Canadian price data. To compare the real-time performance of the Markov switching approach to the benchmark model without regime-switching, we assess their abilities to forecast headline inflation and minimize revisions. We find that the Markov switching model delivers superior predictive accuracy and significantly reduces revisions during periods of substantial inflation changes. Hence, our findings suggest that accounting for time-varying factors and parameters enhances inflation signal accuracy and reduces data requirements, especially following sudden economic shifts.
\\
%\textbf{JEL Classification:} \\
\end{abstract}
\thispagestyle{empty}
%\newpage
%\tableofcontents
%\thispagestyle{empty}

%\setcounter{page}{1}

%%%%%%%%%%%%%%%%%%%%%%%%%%%%%%%%%%%%%%%%%%%%%%%%%%%%%%%%%%%%%%%%%%%%%%%%%%%%%%%%%
%%                           INTRODUCTION                                      %% 
%%%%%%%%%%%%%%%%%%%%%%%%%%%%%%%%%%%%%%%%%%%%%%%%%%%%%%%%%%%%%%%%%%%%%%%%%%%%%%%%%
\linespread{1.5}\selectfont
\newpage
\setcounter{page}{1}
% --------------------------------------
\section{Introduction}
% --------------------------------------
Inflation-targeting central banks use various types of core inflation indicators to asses inflationary pressure. Features that make these indicators useful include being robust to high-frequency volatility from transitory and sector-specific shocks, reliability in minimizing revisions of historical estimates,  forecasting headline inflation in the short to medium-term, and being useful in real-time. The main objective of core inflation indicators are to provide a signal of underlying and overall inflation for a given economy. As a result, indicators with such features can serve as crucial tools for guiding short-term monetary policy decisions. Types of core inflation indicators include: CPI less the most volatile items, Weighted CPI, Median CPI, Trimmed CPI, and other model-based approaches. One of these model based approaches involve using factor models to extract the common component over many individual prices indices. These methods of obtaining a measure of underlying core inflation leverage high dimensional data sets and typically posses the ideal features of a core inflation indicator previously mentioned. For this reason, this paper focuses specifically on measures of underlying core inflation built using factor models. 

These models are used by various central banks to obtain a measure of core inflation. For example, the Bank of Canada (BoC) computes CPI-Common, as detailed in \cite{khan_common_2013} and \cite{khan_comprehensive_2015}. The U.S. Federal Reserve Bank of New York has two versions of their Underlying Inflation Gauge (UIG), as described in \cite{amstad_real_2009} and \cite{amstad_new_2017}, one using price data only while the other complements the same price data with other macro variables. Others include the United Kingdom (see \cite{kapetanios_note_2004}), the Euro area (see \cite{cristadoro_core_2005}), New Zealand (see \cite{giannone_new_2007}; \cite{kirker_what_2010}), and Turkey (see \cite{tekatli_new_2010}). 

However, it is important to acknowledge the inherent challenges associated with these indicators that are built using factor models. Specifically, historical estimates are revised as new data becomes available and there is debate over the number of factors that should be extracted from the data. In most cases, when dealing with inflation, a single factor is assumed to be needed as it is associated with underlying inflation. However, other econometric procedures such as those described in \cite{bai_determining_2002} can also be used to determine the appropriate number of factors. Due to revisions \cite{wynne_core_2008} criticizes the factor model-based approaches for measuring underlying inflation, but acknowledges that if historical revisions were found to be trivial then this criticism would lose power. In the past, revisions were indeed typically negligible, which bolstered confidence in this approach. Yet, with the recent surge in inflation observed across various countries, these revisions have become considerably more concerning. For example, in our study, we find a that the BoC's CPI-Common underestimated underlying inflation by approximately 2.47\% in April 2022 when compared to the estimate obtained when using further data up to December 2022. Such cases have lead some central banks, including the BoC, to reconsider the use of such indicators (see \cite{macklem_whats_2022}).%

In this paper, we address the challenge of large revisions in core inflation estimates observed during changing inflationary regimes by modernizing the factor-based approach using recently proposed econometric procedures. Specifically, we propose two novel core inflation indicators that account for time-varying dynamics and multiple inflation regimes. First, we apply the methods developed by \cite{baltagi_estimating_2021} to detect structural breaks in high-dimensional factor models to develop an underlying core inflation indicator with structural change. Second, we incorporate estimation methods for high-dimensional factor models with Markov switching, as proposed by \cite{urga_estimation_2024}, to develop an underlying core inflation indicator that adapts to regime shifts. Prior research on total and trend inflation has shown compelling evidence for time-varying models (see \cite{stock_has_2002}; \cite{perron_testing_2020}; \cite{amisano_money_2013}; \cite{stock_why_2007}; \cite{stock_core_2016}), supporting our argument that core inflation should also be modeled as subject to multiple regimes.

Both proposed indicators effectively eliminate or reduce historical revisions while retaining the desirable features of core inflation indicators previously mentioned. Moreover, the Markov-switching model proves particularly useful in real-time. Our empirical application using Canadian price data supports the suggested advantages of these indicators. The Markov-switching approach, in particular, demonstrates superior predictive accuracy and significantly reduces revisions during periods of major inflation changes. This is especially relevant for Canada, where our indicators directly address issues with the BoC’s CPI-Common identified by \cite{sullivan_examining_2022} concerning revisions. Specifically, our findings suggest that the CPI-Common with Markov switching, which we call CPI-Common-MS, outperforms the basic CPI-Common in capturing inflation dynamics during periods of significant inflation fluctuation, whether inflation is rising sharply or beginning to stabilize. This effectiveness may stem from the inclusion of data from a past high-inflation period in Canada, specifically between January 1990 and late 1991, prior to the Bank's adoption of inflation targeting. Furthermore, since Canadian price data itself is not revised, it provides an ideal context for evaluating the real-time properties of this indicator. In addition, given this property of the Canadian price data, the CPI-Common with structural change, which we call CPI-Common-SC, is able to completely eliminate revisions for past regimes. However, it can only serve as an offline indicator as, despite its flexibility, real-time detection of breaks is unfortunately not feasible with this method.

As a result, this paper contributes to the literature in four key ways. First, we introduce core inflation indicators that retain the beneficial attributes of traditional core inflation measures but are more robust to abrupt shifts in inflation. They also enhance underlying inflation signals and can serve as a short-term guide for monetary policy. Second, we contribute valuable insights to the ongoing debate over CPI-Common and factor model-based core inflation indicators by proposing avenues for improvement, thereby deepening the debate on core inflation measurement and its implications for effective monetary policy. Third, our CPI-Common-SC and CPI-Common-MS indicators perform particularly well for Canada and present themselves as suitable alternatives to the BoC’s CPI-Common. Finally, our methods can identify dates when underlying inflation, represented as the common factor among various price indices, undergoes regime changes. This has significant implications for inflation-targeting central banks facing elevated inflation levels. Notably, the Markov-switching approach offers the additional benefit of providing policymakers with a probabilistic statement about the state of inflation for a given economy.

The next sections are structured as follows. Section 2 briefly reviews core inflation indicators and outlines how to use a factor model approach to obtain estimates of underlying inflation. Section 3 introduces the core inflation indicators with multiple regimes that we propose here. Section 4 presents the empirical application using Canadian price data and introduces the CPI-Common-SC and CPI-Common-MS indicators that we propose as an alternative to the BoC's CPI-Common. Finally, section 5 provides concluding remarks and some directions for future research.% 

%%%%%%%%%%%%%%%%%%%%%%%%%%%%%%%%%%%%%%%%%%%%%%%%%%%%%%%%%%%%%%%%%%%%%%%%%%%%%%%%%
%%                                   MODELS                                    %%
%%%%%%%%%%%%%%%%%%%%%%%%%%%%%%%%%%%%%%%%%%%%%%%%%%%%%%%%%%%%%%%%%%%%%%%%%%%%%%%%%
% --------------------------------------
\section{Core Inflation}
% --------------------------------------
Inflation-targeting central banks utilize various core inflation indicators to gauge underlying inflation trends, aiming to filter out short-term volatility caused by transitory or sector-specific shocks. The purpose of core inflation measures is to provide a stable, reliable signal of persistent inflationary pressure, which is crucial for guiding monetary policy in the short term. Methodologies to obtain core inflation measures can broadly be categorized into exclusion measures, factor models, filtering approaches, machine learning approaches, and non-parametric procedures, each offering a unique approach to identifying a signal of underlying inflation. The most common approaches used by central banks are exclusion measures and factor models, though machine learning approaches, such as the one proposed in \cite{phil_ml_coreinf}, present an interesting avenue as they are specifically calibrated for optimal forecasting performance.

Exclusion-based indicators, such as CPI excluding energy or other volatile items, are straightforward and aim to eliminate specific components from the Consumer Price Index (CPI) that are known for high volatility. Such measures are simple to interpret but aren't always a reliable measure of underlying inflation when other sectors experience short-term shocks. Another subset within exclusion measures includes trimmed and median CPI indicators. These approaches remove extreme price changes, with the median approach focusing on the middle of the price distribution and trimming techniques excluding certain portions of the distribution tails. As highlighted by \cite{khan_core_2022}, trimmed and median CPI indicators offer robustness by not relying heavily on historical patterns, making them especially useful during periods of economic instability. However, interpretability becomes more nuanced with these measures and by excluding certain items, these exclusion measures might even inadvertently ignore significant inflation trends.

The factor model approach represent a prominent methodology for deriving a signal of underlying inflation and hence building a measure of core inflation. The approach involves estimating a common factor that captures the shared underlying trend among a large set of price indices. Static factor models, such as the Bank of Canada’s CPI-Common, extract these factors under the assumption that the common trend remains constant over time. In general, factor models can often provide better predictions of macroeconomic variables over short- to medium-term horizons (see \cite{stock2002macroeconomic}). For measuring underlying inflation, they have shown to effectively identify persistent inflation drivers, which are less volatile and hence more stable indicators of future inflation, particularly when large datasets are available. However, during recent times of structural changes, we have seen some factor-based methods, like the BoC's CPI-common, be subject to large historical revisions, which is discussed in more detail below. 

For these reasons, policymakers will typically refer to both types of measures as each can shed different insight. However, we believe it is important to improve either type of measure when possible and hence the motivation for introducing an improved methodology for obtaining a core inflation indicator using the factor-based approach in the next section.

First, to formalize core inflation measures, we can assume that headline inflation, $\pi_{t}$, can be expressed as the sum of two unobservable components, a signal of underlying inflation, $\pi_{t}^{*}$, and another term, $\epsilon_{t}$, that combines transitory and idiosyncratic shocks. Hence, we begin with the relationship:
\begin{align}
        \pi_{t} = \pi_{t}^{*} + \epsilon_{t} \label{eq:basic_rela}
\end{align}
To motivate the use of a factor model approach, we define underlying inflation to be a weighted sum of unobserved common factors such that
\begin{align}
    \pi_{t}^{*} = \sum^{r}_{i=1} \omega_{i} f_{t}
\end{align}
where $f_{t}$ is a ($r \times 1$) vector and $\omega_{i}$ are weights, which may be fixed or estimated. Although we can use methods such as \cite{bai_determining_2002} to determine the number of common factors $r$ to use, in this literature it is typically assumed that $r=1$, so that we have a single common component, representing underlying inflation. Hence, we fix $\omega_{1}=1$ and now $f_{t}$ is a scalar that we could estimate from a factor model, for example, using PCA. However, as is common when estimating factor models, we obtain an estimate of the standardized factors due to the identification assumptions needed to identify them. That is, we will instead obtain an estimate of $\tilde{f}_{t}$ from 
\begin{equation}
    x_{t} = \Lambda \tilde{f}_{t} + e_t \label{eq:static_factor_mdl}
\end{equation}
where $x_t = (x_{1,t},\dots, x_{N,t})'$ is a vector of $N$ standardized disaggregated inflation indices and $\Lambda$ is a ($N \times r$) matrix of factor loadings. For core inflation indicators, these typically include disaggregated price or inflation indices that span all sectors of the economy and in some cases, they may even include lags of these price indices, lags of inflation, and other macroeconomic variables. Moreover, the estimate of the common component can further be decomposed as
\begin{equation}
    \hat{\tilde{f}}_{t} =  \tilde{f}_{t} + \tilde{\eta}_{t} 
\end{equation}
where $\tilde{\eta}_{t}$ is some estimation error and $\tilde{f}_{t} = \frac{f_{t} - \mu_{f}}{\sigma_{f}}$. From here, it is easy to see that
\begin{equation}
    f_{t} = \mu_{f} + \sigma_{f}\hat{\tilde{f}}_{t} - \sigma_{f}\tilde{\eta}_{t} 
\end{equation}
and so
\begin{equation}
    \pi_{t}^{*} = \alpha + \beta \hat{\tilde{f}}_{t} + \eta_{t} 
\end{equation}
where $\alpha = \mu_{f}$, $\beta = \sigma_{f}$, and $\eta_{t} = - \sigma_{f}\tilde{\eta}_{t}$. Subbing this back into our original equation, we find that
\begin{align}
        \pi_{t} = \alpha + \beta \hat{\tilde{f}}_{t} + u_{t} \label{eq:pi_common}
\end{align}
where $u_{t} = \eta_{t} + \epsilon_{t}$ and hence, in order to get an estimate of underlying inflation, $\hat{\pi}^{*}_{t} = \hat{\alpha} + \hat{\beta}\hat{\tilde{f}}_{t}$, we can simply estimate the regression equation (\ref{eq:pi_common}) and use the fitted values since, as we can see from equation (\ref{eq:basic_rela}), $\hat{\pi}_t = \hat{\pi}^{*}_{t}$ and all variables in (\ref{eq:pi_common}) are either observable or can be treated as such. It is also worth noting that equation (\ref{eq:pi_common}) can be thought of as a simple factor-augmented model. 
        
An important issue that arises in using this method is that, each time new data becomes available, $\hat{\tilde{f}}_{t}$ is re-estimated and hence $\hat{\pi}^{*}_{t}$ must be re-estimated also, leading to revisions to historical estimates. Notably, with regards to the BoC's CPI-Common, \cite{sullivan_examining_2022} shows that large revisions are typically due to three main sources: i. revisions to the mean of inflation , $\alpha$, ii. revisions to the common factor, $\hat{\tilde{f}}_{t}$, and iii. revisions to the sensitivity of headline inflation to the common factor, $\beta$, with revisions to the latter two being the largest contributors. Notably, the methodology presented below addresses these features directly, which leads to smaller revisions. 

The evaluation of core inflation measures largely depends on: adequately capturing a signal of underlying inflation, minimizing revisions, and reliability in real-time settings. For instance, \cite{khan_comprehensive_2015} outline evaluation methods that examine the reliability of core inflation indicators along these three dimensions. Measuring the ability of core inflation indicators in capturing a signal of underlying inflation is achieved by assessing their ability to forecast headline inflation in the short and medium-term (e.g., up to 12-month out-of-sample). In \cite{khan_uzeda_rodron} a distinction between ``full-information'' measures, which incorporate all available data, and ``real-time'' measures, which rely only on the information available at each historical point in time, are defined and used to further asses real-time performance of these indicators. Evaluating a model’s performance in real-time, as opposed to a full-sample retrospective evaluation, provides a more accurate assessment of its applicability as a short-term guide for monetary policy. However, this can become complicated when price data are revised as is the case with U.S. PCE data (see \cite{FederalReserve2000}; \cite{croushore2008revisions}; \cite{Sinclair2024}). On the other hand, Canadian price data, made available by Statistics Canada, is never revised, making it an excellent setting to properly asses the real-time properties of the core inflation indicators we proposed in the following section. 

% --------------------------------------
\section{Core Inflation with Multiple Regimes}
% --------------------------------------
In this section we begin by introducing the underlying core inflation indicator which accounts for multiple regimes that we propose, under a very general setting. Next, we introduce the two different versions of this indicator that assume different time-varying properties. The first uses a structural change approach and the second uses a Markov switching approach. Structural change and Markov switching methods are both useful when modeling processes that we suspect experience abrupt changes. However, each have their own advantages and disadvantages depending on the application and data availability. For this reason, we do not aim to argue in favor of one of these approaches in general and present both. Throughout, we discuss the features of each indicator and highlight potential shortcomings to be aware of. In doing so, we hope to guide the practitioner in determining the best approach for their application.% 

In general, the indicator we propose is given by
\begin{equation}
    \pi_{t} = \alpha_{j}  + \beta_{j} \hat{\tilde{f}}_{j,t} + \epsilon_t \label{eq:pi_common_regimes}
\end{equation}
where $\pi_{t}$ is still a measure of headline inflation, but now $j$ is used to index the regime and so $j=\{1,2,\dots,M\}$, where M is the number of regimes. Moreover, $\hat{\tilde{f}}_{j,t}$ can be a ($r_{j} \times 1$) vector containing the $r_j$ estimated factors in regime $j$ at time $t$ and $\epsilon_{t}$ is the error process which is assumed to be normally distributed. As previously mentioned, when working with underlying inflation, it is typically assumed that only one common factor is needed as this factor is typically associated with underlying inflation and so we continue to work with $r_j=1$, but this is not necessarily required. The factors are estimated using a time-varying framework in a first step using the standardized series. It is for this reason that, in a second step, we regress total inflation on the estimated factors, treating them as observed, in order to add the mean and standard deviation back and get an indicator that is comparable to other core inflation indicators. We believe that, in general, it may be useful to assume that the mean and the standard deviation of inflation is also time-varying under this setting and as a result, we could assume $\alpha_j$ and $\sigma^2_j$ are also time-varying, in addition to $\beta_j$. To do this, we could simply impose that these parameters change regimes at the same time as the factors estimated in the first step. It is important to note, however, that the dates at which the regimes change are also estimates obtained in the first step. A joint estimation procedure could provide some efficiency gains, but we leave this for future research. 

In the first step, the factors are estimated from the following model
\begin{equation}
    x_t = \Lambda_{j} \tilde{f}_{j,t} + \zeta_t \label{eq:tv_factor_mdl}
\end{equation}
where $x_t = (x_{1,t},\dots, x_{N,t})'$ are as before, but now, $\Lambda_{j} = (\lambda_{j,1},\dots, \lambda_{j,N})'$, where each $\lambda_{j,i}$ can be an $r_j$ dimensional vector, and $\zeta_t = (\zeta_{1,t},\dots, \zeta_{N,t})'$ and assume $\zeta_t \sim \mathcal{N}(\pmb{0}, \Sigma_j)$. Here, we also allow the variance of the factors to be time-varying since
\begin{align*}
    \Sigma_{j} = \Lambda_j\Lambda_j' + \Sigma_{e}
\end{align*}
where $\Sigma_{e}$ may be non-diagonal such that it includes cross-sectional dependence. 

Next, we impose some structure on how the $j$ regimes are determined by considering the structural change or Markov switching approach.%

% --------------------------------------
\subsection{Structural breaks in factor models}
% --------------------------------------
Using a structural change approach has some favorable advantages. It allows us to consider multiple regimes without assuming the same regimes reoccur or that past and future regimes are similar. Specifically, we do not assume anything about the process governing the changes in regime. We only focus on identifying dates when underlying core inflation experiences a structural change and use these break dates to identify the regimes. Hence, when using the structural change approach, we are focused on identifying the $\kappa$ break dates. Here, $\kappa = 1, \dots, m$ where $m+1=M$ is the total number of regimes. The resulting break dates of interest are then $T_{1}, \dots, T_m$ and we use the convention that $T_0=0$ and $T_{m+1}=T$ where $T$ is simply the sample size. 

To estimate these break dates we use the least-squares procedure and dynamic algorithm discussed in \cite{baltagi_estimating_2021}. The estimator and dynamic algorithm for detecting globally optimum break dates discussed by \cite{baltagi_estimating_2021} are very similar in nature to the methods first discussed in \cite{bai_estimating_1998}, \cite{perron_testing_2020} and \cite{bai_computation_2003} for conventional univariate time series. Additionally, in \cite{baltagi_estimating_2021} the authors also propose a series of hypothesis testing procedures that are useful in determining whether there are indeed structural breaks and to determine the appropriate number of breaks. These testing procedures also draw inspiration from the hypothesis testing procedures discussed in \cite{bai_estimating_1998}, which have come to be standard in the econometric literature. It is worthwhile to note that \cite{duan_quasi-maximum_2022} also propose a quasi-maximum likelihood (QML) estimator and a Likelihood ratio based test for detecting and testing breaks in high-dimensional factor models that are similar in style to the methods proposed in \cite{qu_estimating_2007}. However, later in the empirical section we will focus on the least-square estimation method discussed in \cite{baltagi_estimating_2021} and for this reason, we review these procedures now. Specifically, the authors propose the following, sup$F$, test statistic
\begin{align}
    \underset{(\tau_1, \dots, \tau_l) \in \Lambda_{\epsilon}}{\text{sup}} F_{NT}\left(\tau_1,\dots,\tau_l;\frac{\tilde{r}(\tilde{r}+1)}{2}\right)
\end{align}
where $\tilde{r}$ is the number of estimated factors and 
\begin{align}
    F_{NT}\left(\tau_1,\dots,\tau_l;\frac{\tilde{r}(\tilde{r}+1)}{2}\right) = \frac{2}{l\tilde{r}(\tilde{r}+1)}\left[SSNE_0-SSNE(T_1,\dots,T_l)\right]
\end{align}
where
\begin{align*}
    \text{SSNE}(T_{1},\dots, T_{l}) = & \sum^{L+1}_{\kappa=1} \sum^{T_{\kappa}}_{t=T_{\kappa-1}+1} \text{vech}\left(\hat{\tilde{f}}_{t}\hat{\tilde{f}}'_{t} - \frac{1}{T_{\kappa} - T_{\kappa-1}} \sum^{T_{\kappa}}_{t=T_{\kappa-1}+1} \hat{\tilde{f}}_{t}\hat{\tilde{f}}'_{t}\right)' \tilde{\Omega}^{-1} \\
    & \hspace{2.5cm} \text{vech}\left(\hat{\tilde{f}}_{t}\hat{\tilde{f}}'_{t} - \frac{1}{T_{\kappa} - T_{\kappa-1}} \sum^{T_{\kappa}}_{t=T_{\kappa-1}+1} \hat{\tilde{f}}_{t}\hat{\tilde{f}}'_{t}\right)
\end{align*}
\begin{align*}
    \text{SSNE}_0 = & \sum^{T}_{t=1} \text{vech}\left(\hat{\tilde{f}}_{t}\hat{\tilde{f}}'_{t} - \frac{1}{T} \sum^{T}_{t=1} \hat{\tilde{f}}_{t}\hat{\tilde{f}}'_{t}\right)' \tilde{\Omega}^{-1} \text{vech}\left(\hat{\tilde{f}}_{t}\hat{\tilde{f}}'_{t} - \frac{1}{T} \sum^{T}_{t=1} \hat{\tilde{f}}_{t}\hat{\tilde{f}}'_{t}\right)
\end{align*}
and $\tilde{\Omega}$ is a HAC estimate of the long-run covariance matrix. As can be seen, the test statistic is essentially the maximum $F$-statistic over various possible break dates. It will also depend on the parameter $\epsilon$, which determines the minimum length of a regime. As in \cite{bai_estimating_1998}, \cite{baltagi_estimating_2021} propose a UDmax and WDmax test that are used to test up to an upper limit $L$ number of breaks. These are given by
\begin{align}
    \text{UDmax} & = \max_{1\leq l\leq L} \underset{(\tau_1, \dots, \tau_l) \in \Lambda_{\epsilon}}{sup} F_{NT}\left(\tau_1,\dots,\tau_l;\frac{\tilde{r}(\tilde{r}+1)}{2}\right)
\end{align}
\begin{align}
    \text{WDmax} & = \max_{1\leq l\leq L} \frac{c(\nu,\alpha,1)}{c(\nu,\alpha,l)} \underset{(\tau_1, \dots, \tau_l) \in \Lambda_{\epsilon}}{sup} F_{NT}\left(\tau_1,\dots,\tau_l;\frac{\tilde{r}(\tilde{r}+1)}{2}\right)
\end{align}
These tests are meant to detect whether we have breaks at all by considering a null hypothesis of no breaks against an alternative of $\kappa=1$ breaks up to a maximum of $\kappa=L$ breaks. Each time we store the test statistic and then find the maximum over the $L$ test statistics. The UDmax test assumes uniform weights on each test statistic and the WDmax test imposes some weights. For example, the weights described in \cite{baltagi_estimating_2021}, and shown above, are the ratio of the critical values of the test of $0$ versus $1$ break over the critical value for the test of $0$ versus $\kappa$ breaks. In these equations, $\nu = \frac{\tilde{r}(\tilde{r}+1)}{2}$. The authors also propose a sequential $F$($l|l+1$) test that can be used to determine the appropriate number of breaks and is given by
\begin{align}
    F(l|l+1) = SSNE(T_1,\dots,T_l) - \min_{1\leq\iota\leq l+1 \in \Lambda_{\iota,\epsilon}} SSNE(T_1,\dots,T_{\iota-1}, \tau, T_{\iota}, \dots, T_l)
\end{align}
The null distribution of these test procedures have a similar form as those described in \cite{bai_estimating_1998} and \cite{bai_critical_2003} and so, as suggested in \cite{baltagi_estimating_2021}, we can refer to the critical values tabulated there to obtain critical values for these test procedures. Similarly, the process to determine the number of breaks is as originally described in \cite{bai_computation_2003}. That is, we first use the Dmax tests to determine if there are breaks at all. Then we use the sequential approach to determine the number of breaks. The sequential approach keeps the location of the previous breaks under the null fixed and hence do not necessarily find global break dates. This step is only use to determine the number of breaks. Once this number is determined, we estimate the globally optimum break dates using a dynamic algorithm.

It is also important to note that, unlike the conventional structural change procedure discussed in \cite{perron_testing_2020} for univariate time series, the procedures considered here to detecting breaks in factor models are not able to distinguish between breaks in the factor loadings or breaks in the variance of the factors. Hence, more work needs to be done to gain a deeper understanding of what is generating these time-varying dynamics in underlying core inflation. A possible solution includes performing ex-post hypothesis test procedures to determine if the variance of the factors are statistically different across all combinations of the regimes that have been identified. 

Once we obtain estimated break dates $\hat{T}_\kappa = \{\hat{T}_1, \dots, \hat{T}_m\}$, we can estimate the factors in each regime and construct a diagonal matrix partitioned according to the estimated break dates. This gives the following ($T \times q$) matrix 
\begin{align}
    \overline{\widetilde{F}} = \text{diag}(\hat{\tilde{F}}_{1}, \dots, \hat{\tilde{F}}_{m+1})
\end{align}
where each $\hat{\tilde{F}}_{j}$ is a ($(\hat{T}_{j}-\hat{T}_{j-1}) \times r_j$) matrix, which are stacked diagonally and $q$ is the total number of factors across all regimes. In our case, since $r_j=1$, it follows that $q=M$. Similarly, we can generate a matrix of constants partitioned in a similar way. This gives $\overline{I} = \text{diag}(\iota_{T_1}, \dots, \iota_{T_{m+1}-T_m})$. From here, we can express the structural change version of the model given in equation (\ref{eq:pi_common_regimes}) in matrix form as 
\begin{align}
    \Pi = \alpha \overline{I} + \beta \overline{\widetilde{F}} + E \label{sc_mdl}
\end{align}
where now $\alpha$ and $\beta$ are $M \times 1$ and $q \times 1$ vectors. This model can easily estimated by OLS and the fitted values $\hat{\Pi}$ from this regression then give the underlying core inflation indicator with structural change that we propose in this paper. Further, if we wish to account for a time-varying standard deviation of inflation, we could easily estimate the same linear model given in (\ref{eq:pi_common}), within each regime by OLS, and stack the predicted values in a single vector.

The assumption that $\alpha$ is subject to same changes as the common factor(s) can be tested ex-post. However, from (\ref{sc_mdl}), we can see that this model accounts for changes in the mean of inflation, the common factor, and the sensitivity of headline inflation to the common factor, which are all suggested to contribute to large revisions to the CPI-Common, Canada's core inflation indicator, as discussed in \cite{sullivan_examining_2022} and previously mentioned. 

One of the main shortcomings of the structural change approach is that it is an off-line indicator simply because these structural change procedures are off-line detection methods of break dates. Specifically, as discussed in \cite{urga_estimation_2024}, the estimates of the break dates depend on a hyperparameter $\epsilon$. This $\epsilon$ essentially determines the minimum length of a regime. For example, if $T=200$ and $\epsilon=0.1$, then a regime must have at least $20$ observations or more. That is, $T_{\kappa+1} +\epsilon \geq T_{\kappa} \geq T_{\kappa-1} + \epsilon$ $\forall \kappa = 1 \dots, m$. If one is interested in a study where identifying the breaks in underlying core inflation ex-post is of interest, then this poses no issue. However, identifying the most recent break date properly in real-time becomes difficult because we require being quite a few observations, $\epsilon\times T$ observations specifically, into the newest regime to detect it. Further, the flexibility that each regime is unique comes at the cost that, even if we know the most recent break date, we still require being quite a few observations into the newest regime to get a proper estimate of the factors for that regime. 

If on the other hand, we assume that regimes may reoccur, then even if we are only one or two months into a new regime, we can make use of information from the past occurrence of that regime to obtain good estimates of the factors and have a real-time underlying core inflation indicator. The Markov switching approach makes such an assumption and in practice would not even require the previous occurrence of that regime, but its presence may improve estimates nonetheless. For this reason we explore this approach next.
% --------------------------------------
\subsection{Markov switching}
% --------------------------------------
As previously mentioned, the Markov switching approach assumes regimes may reoccur. This is especially favourable if our data spans a large enough period such that this recurrence of regimes is present. Currently, many countries are experiencing higher inflation than in recent history. It is during this increase in inflation that these core inflation indicators built using factor models began to experience much larger revisions. However, most if not all of these countries with inflation-targeting central banks have at some point in the past experienced periods where inflation was much higher and less stable. Specifically before inflation-targeting policies were implemented. As a result, this presents a favourable avenue for the Markov switching approach. 

The difficulty with the Markov switching approach on the other hand, is that we typically need to know the number of regimes we would like to estimate our model with. For conventional Markov switching models (i.e., not applied to factor models), hypothesis testing procedures for determining the number of regimes has been an active area of research for quite some time since \cite{hamilton89} first used these models to model U.S. GNP growth. Noteworthy procedures for testing the number of regimes in these relatively more conventional settings include \cite{hansen92}, \cite{garcia98}, \cite{chp14}, \cite{dufourluger17}, \cite{quzhuo2021likelihood}, and most recently \cite{rodrondufour_mcmstest}. Nevertheless, since estimating high-dimensional factor models with Markov switching is a very recent topic, no procedure is currently available for determining the number of regimes in this setting. However, extending the work of \cite{rodrondufour_mcmstest} for factor models is the subject of ongoing research (e.g., \cite{rodrondufour_mcmsfactortest}) and would provide a good way to determine the number of regimes for underlying core inflation. For now, in the following we will consider models with $M=2$, $M=3$, or $M=4$ regimes and asses their ability to reduce revisions and forecast headline inflation. For example, a case where $M=2$ may be interpreted as having high versus low or stable versus non-stable inflation regimes. 

In order to estimate the factors from a high-dimensional factor model with Markov switching, one could use the procedures discussed in \cite{urga_estimation_2024} or \cite{barigozzi_modelling_2022}. In the first study, the procedure assume we know the number of factors in each regime, though an information criterion is proposed to help with this. In the latter study, the authors also propose a method for estimating the factors that does not assume the number of factors in each regime is known. In general, a Markov switching model with $M$ regimes will also include the one-step transition probabilities, which can be gathered into a transition matrix such as
\begin{align*}
    \textbf{P} & = \begin{bmatrix}
            p_{11} & \dots & p_{M1}\\
            \vdots & \ddots & \vdots \\
            p_{1M} & \dots & p_{MM}
    \end{bmatrix}
\end{align*}
where, for example, $p_{ij} = \text{Pr}(S_{t} = j \mid S_{t-1} = i)$ is the probability of state $j$ following state $i$. The columns of the transition matrix must sum to one to have a well-defined transition matrix  and so $\sum^M_{j=1} p_{ij} = 1$, $\forall i$. Using this transition matrix, \textbf{P}, we can also obtain the ergodic probabilities, $\pi = (\pi_1, \dots, \pi_M)'$, which are given by
\begin{align*}
    \pmb{\pi} = (\mathbf{A}'\mathbf{A})^{-1}\mathbf{A}'\mathbf{e}_{N+1} \hspace{0.25cm}  \&  \hspace{0.25cm} \mathbf{A} = \begin{bmatrix}
        \mathbf{I}_M-\textbf{P}\\
        \pmb{1}'
        \end{bmatrix}
\end{align*}
where $\mathbf{e}_{M+1}$ is the $(M+1)$th column of $\mathbf{I}_{M+1}$. These ergodic probabilities can be understood as representing, in the long-run on average, the proportion of time spent in each regime. These parameters, along with the use of a filtering algorithm, can be used to obtain an estimate of $\text{Pr}(S_{t} = j)$, the probability of being in regime $j$ at time $t$ (see \cite{hamilton1990}; \cite{hamilton1994}; and \cite{urga_estimation_2024}). This estimate can then inform us on the probability of being in any of the $M$ inflation regimes at any date $t$.

The Markov switching version of the core inflation indicator we propose in (\ref{eq:pi_common_regimes}) is given by
\begin{align}
    \pi_{t} = \alpha_{S_t} + \beta_{S_t} \hat{\tilde{f}}_{S_t} + \epsilon_t \label{eq:ms_mdl}
\end{align}
where $\hat{\tilde{f}}_{S_t}$ is governed by the latent Markov process $S_t = \{1,2,\dots,M\}$. Specifically, like the structural change approach described above, here we also require a two-step procedure where we infer $S_t$ based on $\widetilde{f}_{S_t}$ in a first step and impose the assumption that $\alpha$ and $\beta$ are governed by same Markov process. As can be seen, given the time-varying nature of this  core inflation indicator, it is also robust to changes in the mean inflation, changes in the common factor, and changes in the sensitivity of headline inflation to the common factor, which as previously mentioned, have been attributed to being contributors to larger revisions. Further, the underlying core inflation indicator with Markov switching proposed here is useful even as a real-time indicator and provides the probability of being in each regime for each date of our sample. This gives us a probabilistic statement about the state of the inflation in a given economy.

%%%%%%%%%%%%%%%%%%%%%%%%%%%%%%%%%%%%%%%%%%%%%%%%%%%%%%%%%%%%%%%%%%%%%%%%%%%%%%%%%
%%                             EMPIRICAL RESULTS                               %%
%%%%%%%%%%%%%%%%%%%%%%%%%%%%%%%%%%%%%%%%%%%%%%%%%%%%%%%%%%%%%%%%%%%%%%%%%%%%%%%%%

% --------------------------------------
\section{Application to Canadian Price Data}
% --------------------------------------
In this section we present the empirical application with Canadian price data. Specifically, we consider the the BoC's CPI-Common and the price data used to compute it. The CPI-Common is the BoC's measure of core inflation that is built using the factor model approach. It is computed using the 55 components of the CPI. The full list can be found in the appendix of \cite{statistics_canada_bank_2020}. These 55 series are adjusted to remove the effect of changes in indirect taxes, are expressed in year-over-year percentage changes, and are never revised. Here, we may also consider month-over-month percentage changes. However, as discussed in \cite{khan_common_2013}, the use of year-over-year inflation rates eliminates concerns regarding the use of non-seasonally adjusted data and so year-over-year inflation rates remains the main focus. Additionally, seasonally adjusted data is often revised due to revision to the seasonal adjustment factors. Therefore, the use of non-seasonally adjusted Canadian price data is well suited for a proper real-time analysis. Before estimating the factor model, each series is standardized. Price data becomes available in January of 1989 and so year-over-year inflation can only begin as early as January 1990. Hence, for this application we consider data from January 1990 to December 2023. Still, this data goes back far enough to include some observations form the period before the BoC adopted an inflation-targeting policy in 1991. 

\begin{figure}[!tbh]
    \centering
    \caption{BoC Headline CPI \& Core Inflation indicators from January 1990 to December 2024}
    \includegraphics[width=0.90\textwidth]{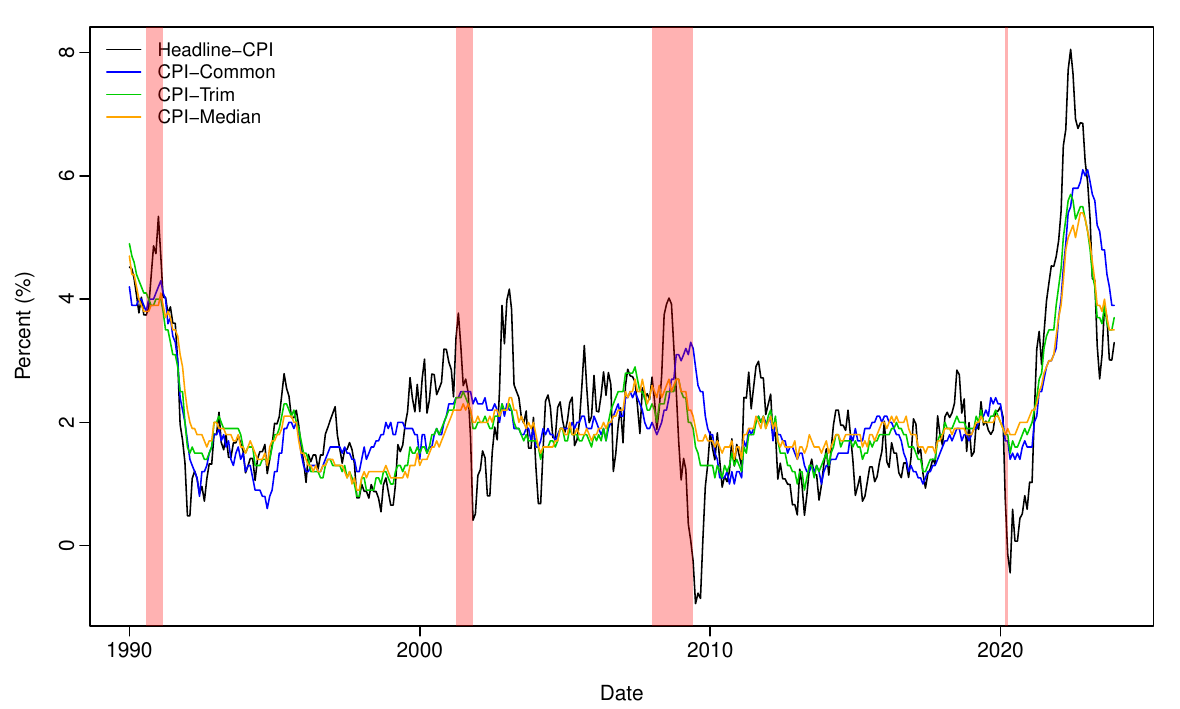}\\
    \label{fig:BoC_inf_measures}
    \parbox[t]{0.95\linewidth}{%
    \scriptsize
    \raggedright 
    \textbf{Notes}: Time series shown here are Headline CPI (black), CPI-Common (blue), the Bank of Canada's time-invariant factor model approach, CPI-Trim (green), and CPI-Median (Orange). Data was obtained from Statistics Canada's website (see \cite{statscan_corecpidata}) and is expressed at a Monthly frequency. Recession dates are shown in red shaded areas and are obtained from NBER (see \cite{nberBCI}).}
\end{figure}

Figure \ref{fig:BoC_inf_measures} shows Canada's Headline CPI index and the BoC's three preferred measures of core inflation, which includes the CPI-Common and two other exclusion-based measures, namely, CPI-Trim and CPI-Median. The red shaded areas highlight recession periods, obtained from National Bureau of Economic Research (NBER), and are meant to provide the interested reader with an idea of the dynamics of these indicators during these recessionary periods. From here, we can see that indeed, the core inflation indicators are less volatile than Headline CPI. We also see that our sample includes two high inflation periods. The first occurring before the BoC imposed the inflation targeting policy and the second occurring more recently. 

\begin{figure}[!tbh]
    \centering
    \caption{CPI-Common April vs. December 2022 revisions}
    \includegraphics[width=0.90\textwidth]{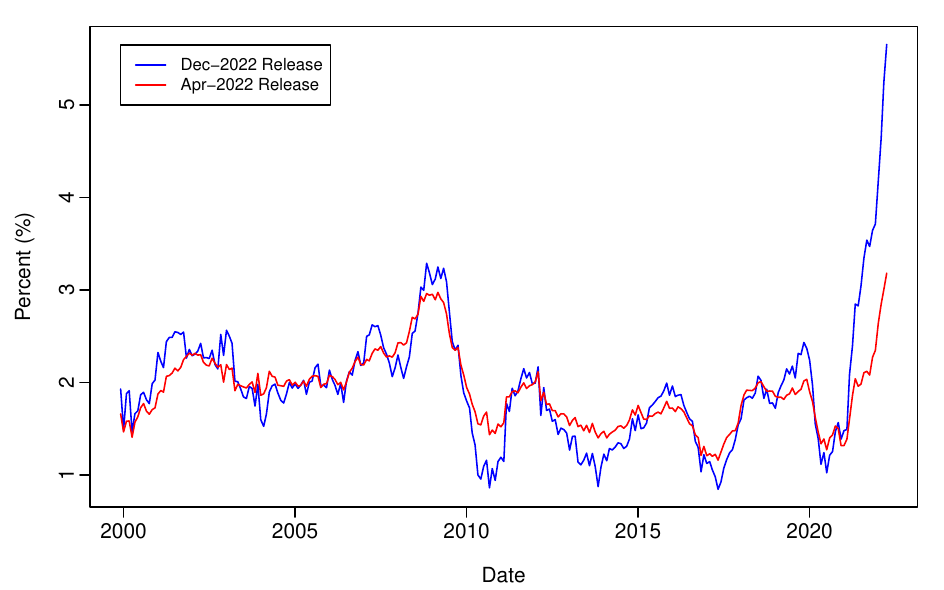}
    \label{fig:worstrev_with_sc}
    \parbox[t]{0.95\linewidth}{%
    \scriptsize
    \raggedright 
    \textbf{Notes}: The red time series line represents the estimates from January 2000 to April 2022, using data only up to April 2022. The blue time series line represents estimates, using the same methodology, from January 200 to April 2022, using data up to December 2022. The difference in estimates for April 2022 is approximately 2.47\%.}
\end{figure}

Using this data we are able to find that, within our sample and in real-time, the worst revision is for the estimate of April 2022 when data up to December 2022 is available. To see this, Figure \ref{fig:worstrev_with_sc} shows the estimates from January 2000 to April 2022 when all data until December 2022 is available (blue) and when only data until April 2022 is available (red). Here, we see that when only data up to April 2022 is available, the estimate for underlying inflation is around 3.18\%. However, with more data from the high inflation period (8 months of additional data), up until December 2022, we find that the estimate for April 2022 should have actually been closer to 5.65\%. This results in a revision of about 2.47\%. Eliminating or reducing such large revisions is the main objective of the alternative CPI-Common-SC and CPI-Common-MS indicators introduced next.

% ------------------
\subsection{CPI-Common-SC} 
% ------------------
To estimate the model given in (\ref{sc_mdl}) using Canadian price data, we must first determine the number of breaks present for this data. Throughout, we assume we only need one common factor as this is how the CPI-Common is currently computed. Using the least-squares estimator of break dates and the sequential test proposed by \cite{baltagi_estimating_2021} and summarized above on Canadian price data, we find evidence of at least $\kappa=6$ breaks when using the shorter minimum regime lengths of $6$ and $12$ months at a significance level of 1\%. If we consider the larger minimum regime length of two years, only $\kappa=4$ breaks are detected. The results of these test procedures are summarized in Table \ref{tab:sc_factor_test_res}.

\begin{table}[!tbh]
\setlength\tabcolsep{8pt}
\caption{Breaks in CPI-Common from January 1990 to December 2023}
\label{tab:sc_factor_test_res}
\begin{center}
\begin{scriptsize}
\begin{tabular}{lcccccccccc}
\hline
\hline
\\[-5pt]
\multicolumn{1}{l}{$\epsilon T$} &    
\multicolumn{2}{l}{Dmax ($M=4$)} & & 
\multicolumn{7}{l}{F($\kappa|\kappa+1$)} \\
\cline{2-3} \cline{5-11}
\\[-5pt]
\multicolumn{1}{c}{} &
\multicolumn{1}{c}{UDmax} &
\multicolumn{1}{c}{WDmax} & & 
\multicolumn{1}{c}{F($1|2$)} &
\multicolumn{1}{c}{F($2|3$)} &
\multicolumn{1}{c}{F($3|4$)} &
\multicolumn{1}{c}{F($4|5$)} &
\multicolumn{1}{c}{F($5|6$)} &
\multicolumn{1}{c}{F($6|7$)} &
\multicolumn{1}{c}{F($7|8$)} \\
\hline
6      & 37.07***   & 37.07***      &   & 37.07***      & 47.31***      & 47.62***      & 12.77 ***     & 8.90***       & 7.89**        &  6.32  \\
12     & 37.07***   & 37.07***      &   & 37.07***      & 47.31***      & 10.49***      & 9.03***       & 7.89***       & 6.50**        &  5.50  \\
24     & 35.89***   & 35.89***      &   & 35.89***      & 44.04***      & 10.84***      & 5.69          & 4.24          & 4.10          &  4.10  \\
\hline
\end{tabular}
\end{scriptsize}
\parbox[t]{0.95\linewidth}{%
\scriptsize
\raggedright 
\textbf{Notes}: The nominal values of $\epsilon=\{0.015,0.03,0.06\}$ and $*$, $**$, and $***$ represent significance at level 1\%, 5\%, and 10\% respectively.}
\end{center}
\end{table}

Specifically, each row considers a different minimum regime length determined by the value of $\epsilon$ times the sample size $T$. The first two columns present the test statistic of the two Dmax tests, namely the UDmax and the WDmax tests. Here we set the maximum number of breaks $m=7$. In this case, they suggest strong evidence in favor of breaks present, rejecting the null hypothesis of no breaks at a $1$\% significance level. Then the next columns present the test statistics for the sequential test. Since we rejected the null of no breaks, we start by consider one break under the null and test against two breaks under the alternative. These test statistics for different values of $\epsilon \times T$ are found in column three. Then, each following column considers a null and an alternative with an additional break. Going up to the last column of the first two rows, we find that when considering a null of seven breaks and comparing this to an alternative of eight breaks, we no longer have enough evidence to reject the null hypothesis of seven breaks at a significance level of $5\%$. The previous column before this one suggests that at a significance level of $1$\% we would fail to reject the null of six breaks in favor of seven. Hence, at a $1$\% significance level, we find that there are at least six breaks (7 regimes) in the common factor when using Canadian price data. 

\begin{figure}[!tbh]
    \centering
    \caption{Headline CPI, CPI-Common, \& CPI-Common-SC from January 1990 to December 2023}
    \includegraphics[width=0.90\textwidth]{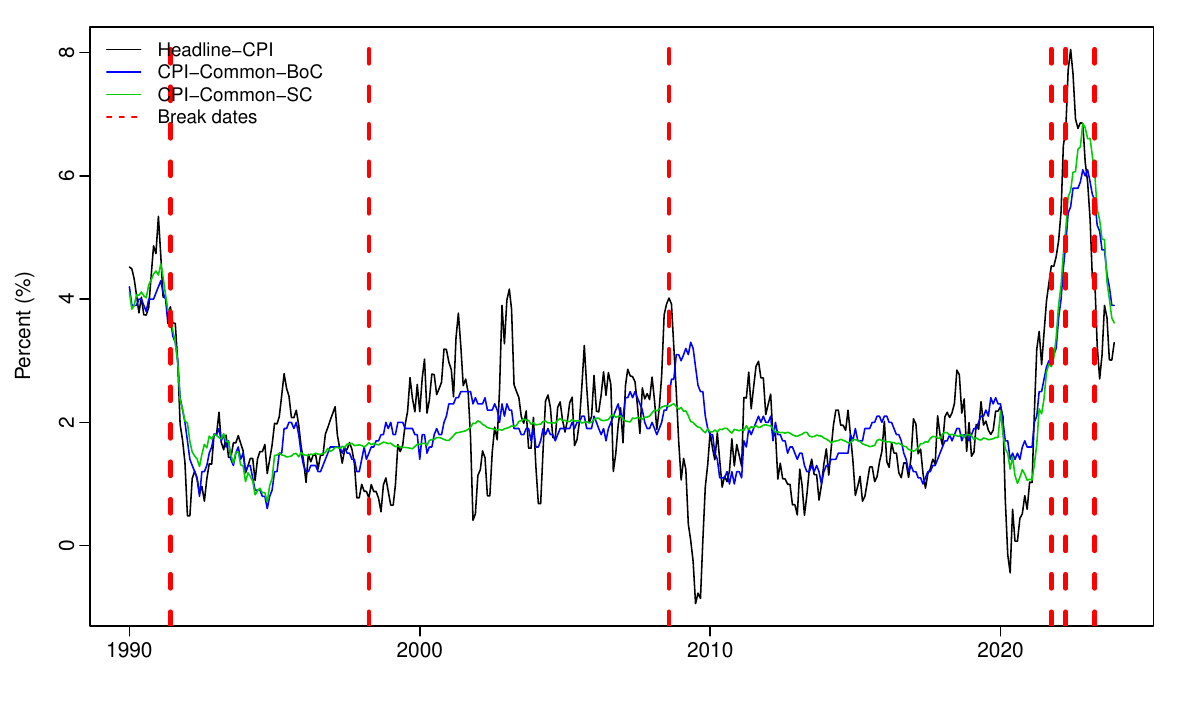}
    \label{fig:cad_sc}
    \parbox[t]{0.95\linewidth}{%
    \scriptsize
    \raggedright 
    \textbf{Notes}: The time series include Headline CPI (black), CPI-Common (blue) - the Bank of Canada’s time-invariant factor model - and CPI-Common-SC, which incorporates the structural break approach proposed here. Red-dotted vertical lines indicate estimated break dates: June 1991, April 1998, August 2008, October 2021, April 2022, and April 2023.}
\end{figure}

These breaks, and the regimes they identify, can be seen in Figure \ref{fig:cad_sc} (red dashed lines). The first break occurs around the time the BoC adopted their inflation-control target. The next two breaks identify three regimes: one where inflation begins to stabilize following inflation targeting, one that begins in the late 1990s and ends around the Great recession, which appears to be more volatile, and another that ends before the recent inflation surge and that appears to be more stable than the regimes that surround it. The last three breaks identify three more regimes: one where inflation is rising, another where inflation is normalizing, and a final regime where, for now, inflation appears to be more stable.
Figure \ref{fig:cad_sc} also shows Headline CPI (black), the BoC's CPI-Common with no structural breaks (blue), and also includes the CPI-Common-SC proposed here (green). Here, we can also see that the proposed CPI-Common-SC is less volatile throughout the sample when compare to the CPI-Common. That is, we find that for the basic CPI-Common, including high inflation periods introduces more volatility to historical estimates, even during the low inflation regimes, while the CPI-Common-SC is robust to this. 

\begin{figure}[!tbh]
    \centering
    \caption{April vs. December 2022 revisions with and without SC}
    \includegraphics[width=0.90\textwidth]{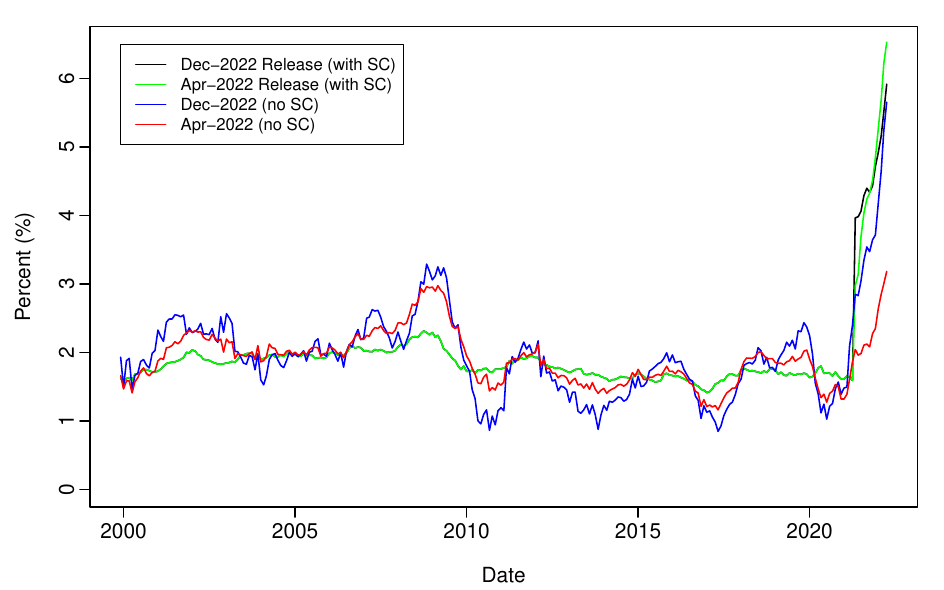}
    \label{fig:worstrev_with_without_sc}
    \parbox[t]{0.95\linewidth}{%
    \scriptsize
    \raggedright 
    \textbf{Notes}: The red line shows time-invariant estimates (benchmark) from January 2000 to April 2022, based on data available up to April 2022. The blue line shows revised benchmark estimates for the same period using data up to December 2022. The green line shows estimates using the structural change approach, up to April 2022, and the black line revised estimates for same period using data up to December 2022. For benchmark model, April 2022 revision is 2.47\% upward and for structural change approach, April 2022 revision is 0.61\% downward .}
\end{figure}

If we revisit the worst revision found in the sample for the basic CPI-Common, now considering our structural change approach we find that the revision for April 2022 is now only about $0.67$\%, which is much smaller than the $2.47$\% when we don't consider structural change. Further, for these two samples, the revisions leading up to April 2022 with the structural change approach are also much smaller. This can be seen in Figure \ref{fig:worstrev_with_without_sc}. 
In addition, for the previous regimes which have ended, this figure also shows that the green and black line are completely overlapping because revisions have now been completely eliminated for those regimes. This is because there are no revision to the price data used to compute the factors and since that regime has ended the estimates of the factors for that regime remain fixed. Figure \ref{fig:worstrevdiff_with_sc} in the appendix show the same information as Figure \ref{fig:worstrev_with_without_sc} but in a slightly different way. Specifically, we show the difference between the blue and red line (the basic CPI-Common) in red and the difference between the black and green line (CPI-Common-SC) in blue. Here, we can see that the difference is flat for the previous regimes. For the CPI-Common-SC, we do not perform a revision study in real time since we understand that this approach is limited in being a off-line procedure. 

% ------------------
\subsection{CPI-Common-MS} 
% ------------------
In this section, we consider four variations of the Markov switching model. Our benchmark model has $M=1$ regime, meaning it does not involve Markov switching and is thus equivalent to the BoC's CPI-Common. This yields estimates denoted by $\hat{\pi}^{M1}_t$. Additionally, we consider three other models: one with $M=2$ regimes, one with $M=3$ regimes, and another with $M=4$ regimes, denoted by $\hat{\pi}^{M2}_t$, $\hat{\pi}^{M3}_t$, and $\hat{\pi}^{M4}_t$, respectively. First, we evaluate these models based on their real-time performance and then summarize the preferred model. To do this, we assess their performance based on real-time versus full-information evaluations, following the definitions in \cite{khan_uzeda_rodron}, and examine their ability to forecast headline inflation in real-time. We also present charts to visually assess their real-time performance, as in \cite{khan_common_2013}.
        
To begin, we review the definitions of real-time and full-information estimates introduced in \cite{khan_uzeda_rodron}, which we use to evaluate the performance of each Markov switching indicator. This approach provides insight into the real-time performance of each measure. Full-information estimates are derived using all the data available in the sample. Thus, we define $\hat{\pi}_{t}^{m,f}=E[\pi_t^{m}|\mathcal{I}_{T}]$, where the superscript $f$ indicates full-information and $\mathcal{I}_{T}$ represents the information set containing the complete sample. For real-time estimates, however, the value $\hat{\pi}_{t}^{m}$ is calculated using only data up to time $t$, without revising prior estimates; only the latest estimate is retained. Thus, we define $\hat{\pi}_{t}^{m,r}=E[\pi_{t}^{m}|\mathcal{I}_{t}]$, where the superscript $r$ denotes real-time, and $\mathcal{I}_{t}$ includes only the information available up to time $t$. Throughout, we use $m={M1, M2, M3, M4}$ to denote the model variations. Since the sample grows over time, even full-information estimates may change, describing only the full information available at a given point. For this reason, we consider five sample periods: a pre-COVID sample (January 2000 - December 2019), a rising inflation sample (January 2020 - December 2022), an inflation normalization sample (January 2023 - December 2023), a post-COVID sample (January 2020 - December 2023), and the full sample (January 2000 - December 2023). In each period, full-information estimates contain data up to the sample’s endpoint. For example, in the rising inflation sample, we compare the performance of our real-time Markov switching indicators as inflation begins to rise with a version that has more information about this period of high inflation. During this time, as seen with the BoC’s CPI-Common and previously discussed, significant revisions occurred as more data on high inflation became available. This exercise aims to illustrate how our measures would perform in such settings, relative to estimates that incorporate additional information about the newest regime. 

For each sample and model, we calculate the Mean Squared Difference (MSD) between the real-time and full-information estimates, using both year-over-year (Y-o-Y) and month-over-month (M-o-M) inflation. Table \ref{tab:real_vs_fullinfo} below presents the results of this analysis.
\begin{table}[!tbh]
\setlength\tabcolsep{6.5pt}
\caption{Root Mean Squared Difference of real-time vs. full-information for each model}
\label{tab:real_vs_fullinfo}
\begin{center}
\begin{scriptsize}
\begin{tabular}{lccccccccccccccc}
\hline\hline
 & \multicolumn{2}{c}{Pre-COVID} & & \multicolumn{2}{c}{Rising Inflation} & & \multicolumn{2}{c}{Inflation Norm.} & & \multicolumn{2}{c}{Post-COVID} & &\multicolumn{2}{c}{Full-Sample} \bigstrut \\ 
\hhline{~ - - ~ - - ~ - - ~ - - ~ - -}
                   & Y-o-Y & M-o-M & & Y-o-Y & M-o-M & & Y-o-Y & M-o-M & & Y-o-Y & M-o-M & & Y-o-Y & M-o-M \bigstrut \\
\hline
$\hat{\pi}^{M1}_t$ & 0.010 & 0.009 & & 1.427 & 0.005 & & 0.042 & 0.002 & & 0.505 & 0.006 & & 0.141 & 0.008 \\
$\hat{\pi}^{M2}_t$ & 0.121 & 0.007 & & 2.249 & 0.023 & & 0.009 & 0.000 & & 0.835 & 0.016 & & 0.238 & 0.015 \\
$\hat{\pi}^{M3}_t$ & 0.174 & 0.012 & & 0.775 & 0.029 & & 0.011 & 0.002 & & 0.158 & 0.030 & & 0.167 & 0.014 \\
$\hat{\pi}^{M4}_t$ & 0.162 & 0.013 & & 0.624 & 0.017 & & 0.544 & 0.019 & & 1.165 & 0.037 & & 0.316 & 0.014 \\
\hline\hline
\end{tabular}%     
\end{scriptsize}
\parbox[t]{0.95\linewidth}{%
\scriptsize
\raggedright 
\textbf{Notes}: Y-o-Y is Year-over-Year while M-o-M is Month-over-Month inflation rates. In this table, we compare real-time estimates against the full information estimates for each model. That is, we use the difference $\hat{\pi}_t^{m,r} - \hat{\pi}_t^{m,f}$ for each model $m=\{M1, M2, M3, M4\}$ and for each sample. Values shown are root mean squared differences. The Pre-COVID sample ranges from 1990-01 to 2019-12, Rising Inflation sample ranges from 2020-01 to 2022-12, Inflation Normalization sample ranges from 2023-01 to 2023-12, Post-COVID sample ranges from 2020-01 to 2023-12, and Full-Sample ranges from 1990-01 to 2023-12.}
\end{center}
\end{table}
Across all cases, the MSD values are generally low. The highest MSDs are observed for the $\hat{\pi}^{M1}_t$ and $\hat{\pi}^{M2}_t$ indicators, particularly during the period when inflation began to rise and when examining year-over-year inflation. In contrast, the $\hat{\pi}^{M3}_t$ and $\hat{\pi}^{M4}_t$ indicators perform significantly better in this period. As inflation begins to normalize, $\hat{\pi}^{M3}_t$ and $\hat{\pi}^{M2}_t$ outperform the others. For the post-COVID period, $\hat{\pi}^{M3}_t$ maintains the best performance overall. Considering the entire sample, both $\hat{\pi}^{M3}_t$ and $\hat{\pi}^{M1}_t$ show the lowest MSD values. Therefore, over all samples, the $\hat{\pi}^{M3}_t$ indicator generally outperforms the others or performs comparably well. Table \ref{tab:real_vs_fullinfo_mad} in the appendix shows the same results when using the Mean Absolute Difference (MAD) and the results presented there are consistent with the ones shown here. 

Next, we assess the performance of our Markov switching indicators in forecasting headline inflation in real time. To do this, we estimate $\hat{\pi}^m_t$ using an expanding window approach, as if in real time, and then forecast headline inflation from 1-month ahead to 12-months ahead over the post-COVID period from January 2020 to December 2023. This produces a forecasting sample, at each horizon, of $T_{f}=48$ observations. Additionally, we examine forecasting over an extended period that includes the Great Recession, from January 2007 to December 2023. In this case, we obtain a forecasting sample, at each horizon, of $T_{f}=204$. To evaluate forecast accuracy, we use the Root Mean Squared Forecasting Error (RMSFE), defined as:
\begin{align}
    \text{RMSFE}^m_h = \sqrt{\frac{\sum^{T_f}_{i} (\hat{\pi}^m_t - \pi^{\text{HCPI}}_{t+h})^2}{T_f}}
\end{align}
To jointly evaluate all models, we apply the Model Confidence Set (MCS) approach proposed by \cite{Hansen-Lunde-Nason2011}. This is a sequential test procedure that can be used to evaluate the forecasting performance of all models and determine a set of models, $\hat{\mathcal{M}}_{(1-\alpha)\%}$, that will contain the best model with a given level of confidence. A higher value of $\alpha$ enforces a stricter elimination rule, resulting in fewer models in $\hat{\mathcal{M}}_{(1-\alpha)\%}$. 

Table \ref{tab:forecasting} presents the results of this forecasting exercise where the $T_{max}$ test statistic and a level of significant $\alpha=0.25$ was used to determine the MCS. Models included in the set $\hat{\mathcal{M}}_{75\%}$ are highlighted in bold for their RMSFE values, with the lowest RMSFE value in red. The values in parenthesis are the p-values of the tests. The p-values of the models not in the model confidence set are obtained from the previous step, before the model was eliminated from $\hat{\mathcal{M}}_{75\%}$. These results were obtained using the R package \textit{MCS} by \cite{CataniaBernardiMCS} and described in \cite{bernardi2014modelconfidencesetpackage}, with modifications to retain the p-values of eliminated models. The results show that the $\hat{\pi}^{M3}_t$ and $\hat{\pi}^{M4}_t$ indicators consistently belong to the MCS when forecasting Y-o-Y inflation. Although the $\hat{\pi}^{M4}_t$ indicator has the lowest RMSFE, it performs similarly to $\hat{\pi}^{M3}_t$. Figures \ref{fig:realtime_fcast_2020} and \ref{fig:realtime_fcast_2007} in the appendix provide a graphical summary of this information, confirming that the RMSFE of the $\hat{\pi}^{M3}_t$ and $\hat{\pi}^{M4}_t$ indicators remain very close across forecast horizons and for both forecasting samples. Table \ref{tab:forecasting_tr_M75} in the appendix presents results when using the $T_{R}$ test statistic instead of $T_{max}$ and show that these results do not depend on the test statistic used.

\begin{table}[!tbh]
\setlength\tabcolsep{7pt}
\caption{Forecasting Headline CPI $\pi^{\text{HCPI}}_{t+h}$}
\label{tab:forecasting}
\begin{center}
\begin{scriptsize}
\begin{tabular}{lccccccccccc}
\hline\hline
\multirow{2}{*}{Models} & \multicolumn{5}{c}{Year-over-Year}  & & \multicolumn{5}{c}{Month-over-Month} \bigstrut\\
\cline{2-6} \cline{8-12}
 & $h=1$ &  $h=3$ &  $h=6$ &  $h=9$ &  $h=12$ &   & $h=1$ &  $h=3$ &  $h=6$ &  $h=9$ &  $h=12$  \bigstrut \\  
\hline
& \multicolumn{11}{c}{$h$-step ahead from 2020-01 to 2023-12 ($T_f=48$)} \bigstrut\\
\hline
\multirow{2}{*}{$\hat{\pi}^{M1}_t$}  &                           1.823  &                           2.157  &                           2.624  &                           2.914  &                  \textbf{3.093}   &   &                  \textbf{0.601}   &                           0.653   &                  \textbf{0.676}   &                  \textbf{0.596}   &                           0.520  \\
                                       &                         (0.023)  &                         (0.011)  &                         (0.054)  &                         (0.182)  &                         (0.478)   &   &                         (0.915)   &                         (0.000)   &                         (0.665)   &                         (0.605)   &                         (0.223)  \\
\multirow{2}{*}{$\hat{\pi}^{M2}_t$}  &                           1.626  &                           2.041  &                           2.576  &                           2.973  &                           3.232   &   & \textbf{\textcolor{red}{0.591}}   &                  \textbf{0.597}   &                  \textbf{0.651}   &                  \textbf{0.580}   &                  \textbf{0.492}  \\
                                       &                         (0.019)  &                         (0.022)  &                         (0.078)  &                         (0.094)  &                         (0.024)   &   &                         (1.000)   &                         (1.000)   &                         (1.000)   &                         (1.000)   &                         (1.000)  \\
\multirow{2}{*}{$\hat{\pi}^{M3}_t$}  &                  \textbf{1.163}  &                  \textbf{1.542}  &                  \textbf{2.108}  &                  \textbf{2.592}  &                  \textbf{2.978}   &   &                  \textbf{0.594}   & \textbf{\textcolor{red}{0.594}}   &                  \textbf{0.653}   &                  \textbf{0.594}   &                  \textbf{0.493}  \\
                                       &                         (0.289)  &                         (0.250)  &                         (0.302)  &                         (0.280)  &                         (1.000)   &   &                         (1.000)   &                         (1.000)   &                         (1.000)   &                         (0.560)   &                         (0.987)  \\
\multirow{2}{*}{$\hat{\pi}^{M4}_t$}  & \textbf{\textcolor{red}{1.130}}  & \textbf{\textcolor{red}{1.509}}  & \textbf{\textcolor{red}{2.073}}  & \textbf{\textcolor{red}{2.545}}  & \textbf{\textcolor{red}{2.923}}   &   &                  \textbf{0.596}   &                  \textbf{0.610}   & \textbf{\textcolor{red}{0.646}}   & \textbf{\textcolor{red}{0.555}}   & \textbf{\textcolor{red}{0.491}}  \\
                                       &                         (1.000)  &                         (1.000)  &                         (1.000)  &                         (1.000)  &                         (1.000)   &   &                         (1.000)   &                         (0.546)   &                         (1.000)   &                         (1.000)   &                         (1.000)  \\
\hline
& \multicolumn{11}{c}{$h$-step ahead from 2007-01 to 2023-12 ($T_f=204$)} \bigstrut\\
\hline
\multirow{2}{*}{$\hat{\pi}^{M1}_t$}  &                           1.176  &                           1.328  &                           1.537  &                  \textbf{1.667}  &                  \textbf{1.742}   &   &                           0.471   &                           0.504   &                  \textbf{0.502}   &                  \textbf{0.479}   &                  \textbf{0.386}  \\
                                       &                         (0.082)  &                         (0.092)  &                         (0.136)  &                         (0.266)  &                         (0.908)   &   &                         (0.141)   &                         (0.145)   &                         (0.256)   &                         (0.354)   &                         (0.898)  \\
\multirow{2}{*}{$\hat{\pi}^{M2}_t$}  &                           1.053  &                           1.245  &                           1.498  &                           1.685  &                  \textbf{1.796}   &   & \textbf{\textcolor{red}{0.452}}   &                  \textbf{0.489}   & \textbf{\textcolor{red}{0.481}}   &                  \textbf{0.475}   & \textbf{\textcolor{red}{0.379}}  \\
                                       &                         (0.180)  &                         (0.173)  &                         (0.211)  &                         (0.231)  &                         (0.287)   &   &                         (1.000)   &                         (0.364)   &                         (1.000)   &                         (0.256)   &                         (1.000)  \\
\multirow{2}{*}{$\hat{\pi}^{M3}_t$}  &                  \textbf{0.892}  &                  \textbf{1.062}  &                  \textbf{1.320}  &                  \textbf{1.535}  &                  \textbf{1.697}   &   &                  \textbf{0.456}   &                           0.496   &                  \textbf{0.490}   &                           0.484   &                  \textbf{0.381}  \\
                                       &                         (0.098)  &                         (0.078)  &                         (0.059)  &                         (1.000)  &                         (1.000)   &   &                         (0.944)   &                         (0.114)   &                         (0.999)   &                         (0.046)   &                         (1.000)  \\
\multirow{2}{*}{$\hat{\pi}^{M4}_t$}  & \textbf{\textcolor{red}{0.833}}  & \textbf{\textcolor{red}{1.012}}  & \textbf{\textcolor{red}{1.280}}  & \textbf{\textcolor{red}{1.502}}  & \textbf{\textcolor{red}{1.673}}   &   &                  \textbf{0.459}   & \textbf{\textcolor{red}{0.480}}   &                  \textbf{0.485}   & \textbf{\textcolor{red}{0.458}}   &                  \textbf{0.383}  \\
                                       &                         (1.000)  &                         (1.000)  &                         (1.000)  &                         (1.000)  &                         (1.000)   &   &                         (0.400)   &                         (1.000)   &                         (1.000)   &                         (1.000)   &                         (0.999)  \\
\hline\hline
\end{tabular}%     
\end{scriptsize}
\parbox[t]{0.95\linewidth}{%
\scriptsize
\raggedright 
\textbf{Notes}: Reported values are the root mean squared forecast error of model $m$ at horizon $h$ (i.e., $\text{RMSFE}^m_h$). Estimation of $\hat{\pi}^{m}_t$ is performed using an expanding window with information at time $t$ only (real-time). Using the $T_{max}$ test of \cite{Hansen-Lunde-Nason2011} with $\alpha=0.25$. Values in bold highlight models that belong to $\hat{\mathcal{M}}_{75\%}$ and values in \textcolor{red}{red} highlight those that belong to the MCS and have lowest RMSFE. The values in parenthesis are the associated p-values of the MCS test using the sequential algorithm. The p-values of the eliminated models are those from the previous iteration when model was eliminated.}
\end{center}
\end{table} 

Finally, Figure \ref{fig:realtime_rev_1990} in the appendix provides a visual evaluation of revisions for each indicator in a comprehensive way. Following \cite{khan_common_2013}, each panel in this figure displays stacked estimates from each data release within the same chart, for each indicator. Specifically, when looking at each chart, each line represents all estimates calculated using data from the beginning of the sample up to date $t$, including only information available up to time $t$. Each time a new month of data is released, a new line is added, reflecting revisions of past estimates. In the real-time versus full-information analysis discussed above, $\hat{\pi}^{m,r}_{t}$ is a vector containing the last estimate of each line, while $\hat{\pi}^{m,f}_{t}$ is the single line with corresponding end-date. Here, for a given date $t$, a larger spread across all lines suggests more substantial revisions, while periods where the lines are close together indicate fewer revisions. Figure \ref{fig:realtime_rev_2020} displays the same information but starts in January 2020 for easier visual inspection. We observe that when $M=1$ (the basic CPI-Common), there is a large spread when inflation starts to rise. A similar pattern occurs when $M=2$, though the spread during this period decreases when using $M=3$, which supports the findings presented in Table \ref{tab:real_vs_fullinfo}.

\begin{figure}[!tbh]
    \centering
    \caption{Headline CPI, CPI-Common, \& CPI-Common-MS from January 1990 to December 2023}
    \includegraphics[width=0.90\textwidth]{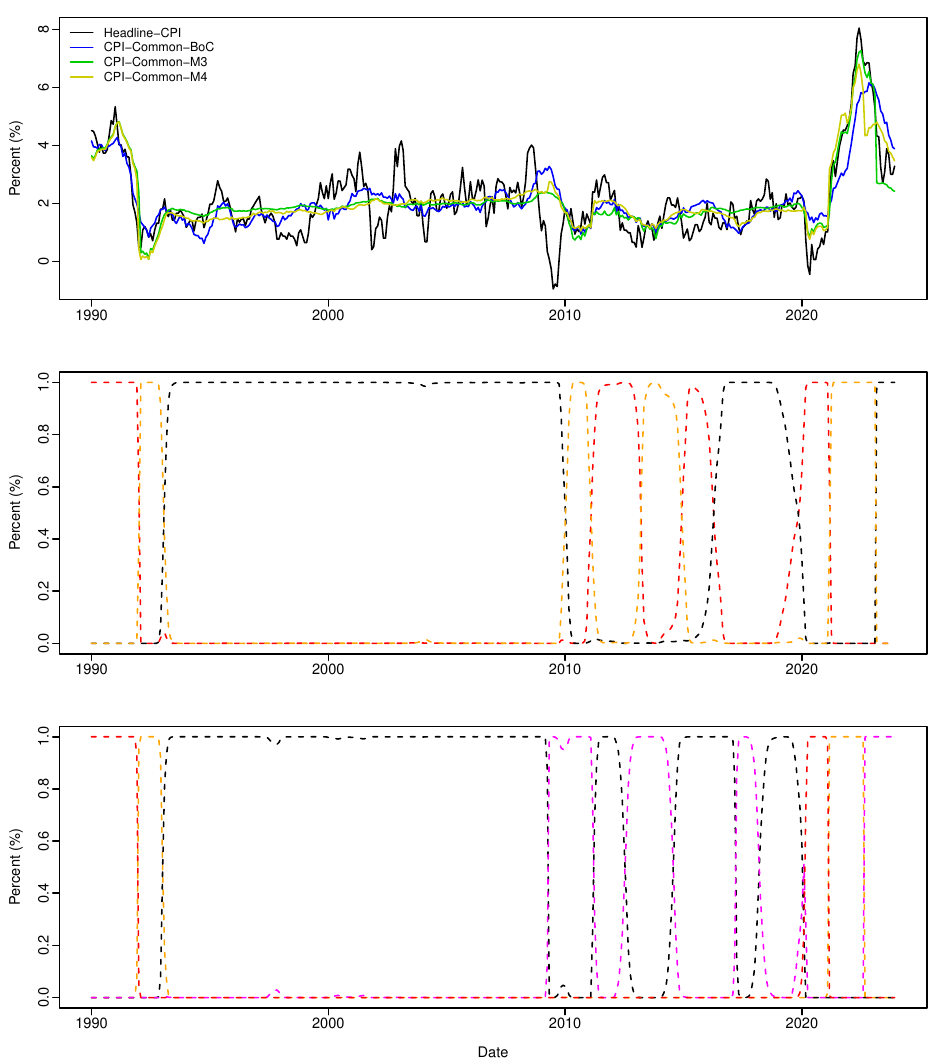}
    \label{fig:indicators_m3_m4_smooth_probs}
    \parbox[t]{0.95\linewidth}{%
    \scriptsize
    \raggedright 
    \textbf{Notes}: Top Chart: The time series include Headline CPI (black), CPI-Common (blue, the Bank of Canada’s time-invariant factor model), CPI-Common-M3 (green, Markov switching with $M=3$ regimes), and CPI-Common-M4 (yellow, Markov switching with $M=4$ regimes). Middle Chart: Time series show estimated smoothed probabilities for regime 1 (red), regime 2 (black), and regime 3 (orange) associated to CPI-Common-M3. Bottom Chart: Time series show estimated smoothed probabilities for regime 1 (red), regime 2 (black), regime 3 (orange), and regime 4 (magenta) associated to CPI-Common-M4.}
\end{figure}

These real-time performance results indicate that, in terms of minimizing revisions, the model with $M=3$ regimes is preferred. Conversely, for forecasting headline inflation in real time, the model with $M=4$ regimes is slightly preferred, though its performance is similar to the model with $M=3$ regimes. Figure \ref{fig:indicators_m3_m4_smooth_probs} presents a graph of headline inflation, the BoC’s CPI-Common (the model with $M=1$ and our benchmark), CPI-Common-MS3 (our proposed indicator with $M=3$ regimes), and CPI-Common-MS4 (our proposed indicator with $M=4$ regimes). We observe that both Markov switching indicators are more stable throughout most of the sample compared to the benchmark CPI-Common, thereby providing a more stable signal of underlying inflation. Much of the basic CPI-Common’s variability arises from the inclusion of high inflationary periods in its calculation, even for periods that don't correspond to high inflation regimes. However, in recent months, CPI-Common-MS3 and CPI-Common-MS4 differ in the inflationary signals they suggest. Specifically, CPI-Common-MS4 indicates that inflation began to normalize sooner, while CPI-Common-MS3 presents an even more optimistic outlook, suggesting that underlying inflation is much closer to the BoC’s target by the end of the sample period.
Figure \ref{fig:indicators_m3_m4_smooth_probs} also displays the estimated and smoothed regime probabilities over time. These probabilities suggest a prolonged regime starting around the time inflation began to stabilize following the adoption of an inflation-targeting policy, lasting until the onset of the Great Recession. In both models, the period between the Great Recession and the recent inflation surge is marked by frequent regime shifts. Notably, in the case of CPI-Common-MS4, the regime probabilities indicate that we are currently in a regime similar to that experienced during the Great Recession, possibly signaling that we are still in a regime of greater uncertainty. In contrast, the smoothed probabilities related to the CPI-Common-MS3 indicator also suggest a return to a more stable regime, once more highlighting a potentially more optimistic view of the current inflationary environment.

In this context, it would be ideal to have a valid hypothesis testing procedure to determine which model, CPI-Common-MS3 or CPI-Common-MS4, sufficiently represents the dynamics in the data. However, such a procedure is not currently available, though it is the subject of ongoing research (see \cite{rodrondufour_mcmsfactortest}). A test like this could be valuable for justifying the simpler CPI-Common-MS3 model. Moreover, it could be applied in real-time, as it is likely that, while both the CPI-Common-MS3 and CPI-Common-MS4 models are suitable for the full sample, a model with $M=2$ regimes might have been sufficient for shorter sample, for example ending before the Great recession. This could also explain the relatively poorer performance of these models during the pre-COVID period, as seen in Tables \ref{tab:real_vs_fullinfo} and \ref{tab:real_vs_fullinfo_mad}.

\begin{figure}[!tbh]
    \centering
    \caption{Correlations of CPI components in full-sample (left) and by regime (right)}
    \includegraphics[width=0.90\textwidth]{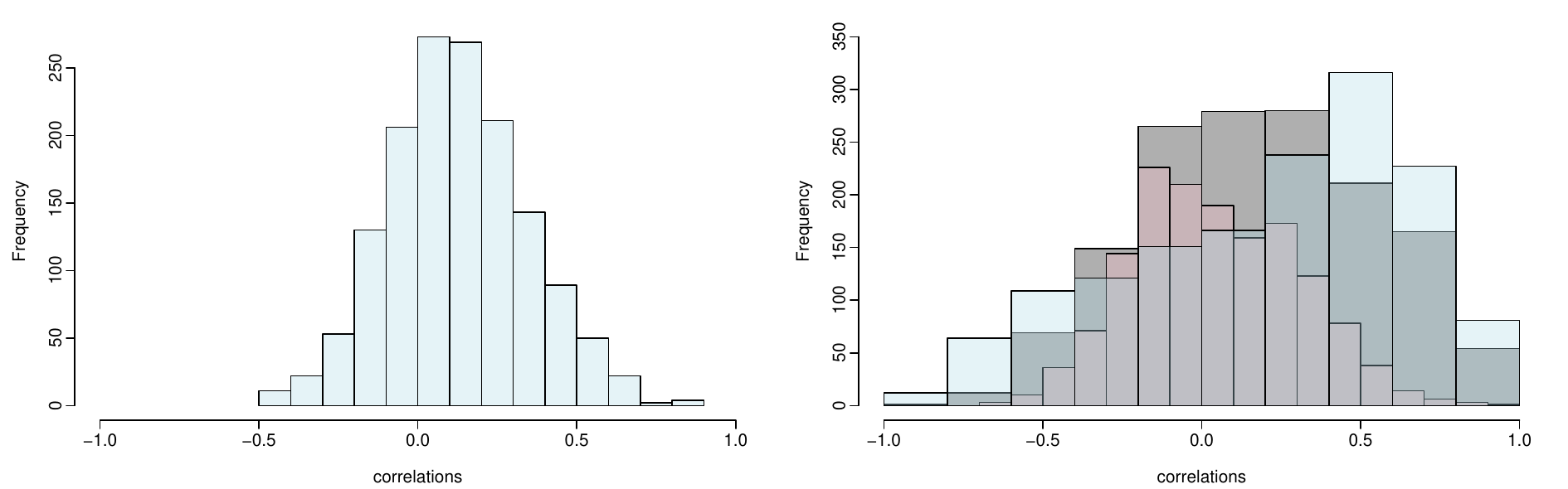}
    \label{fig:cpicommonms34}
    \parbox[t]{0.95\linewidth}{%
    \scriptsize
    \raggedright 
    \textbf{Notes}: Left Chart: histograms of correlations between all 55 CPI components over full sample from January 1990 to December 2023. Right Chart: histogram of correlations between all 55 CPI components within each of the $M=3$ regimes.}
\end{figure}

Nevertheless, given the better performance of CPI-Common-MS3 in the revision analysis, we focus on summarizing this more parsimonious model. Since the factor model approach utilizes information about how components vary together, Figure \ref{fig:cpicommonms34} shows the correlation of the 55 components of the CPI over the full sample on the left, and the correlation of these components within each regime on the right. Specifically, on the right hand side, we see that indeed the distribution of correlations of the components differ across regimes and that they especially differ from the distribution of correlations in the full-sample. This further supports the notion that these three regimes are indeed markedly distinct.

As a result of estimating the Markov switching factor model, we also obtain estimates of the transition probabilities, as well as the limiting (ergodic) probabilities for each regime. These are presented in equation (\ref{eq:transprob}). In the middle chart of Figure \ref{fig:indicators_m3_m4_smooth_probs}, the first regime is represented in red, the second in black, and the third in orange. 
\begin{align}
    \textbf{P} & = 
    \begin{bmatrix}
            0.950 & 0.004 & 0.030\\
            0.012 & 0.992 & 0.030\\
            0.038 & 0.004 & 0.940
    \end{bmatrix} 
    &  
    \pi & = 
    \begin{bmatrix}
            0.144 & 0.716 & 0.14
    \end{bmatrix} 
    \label{eq:transprob}
\end{align}
Combined with the information provided by (\ref{eq:transprob}), the results indicate that the two less stable regimes are, in fact, quite persistent, with high probabilities of transitioning back to the same regime. However, the stable, low-inflation regime remains the most persistent, given its longer duration over the sample period. This persistence is reflected in the ergodic probabilities, which indicate the long-run average time spent in each regime.

\begin{figure}[!tbh]
    \centering
    \caption{Headline CPI \& CPI-Common from January 1990 to December 2023}
    \includegraphics[width=0.90\textwidth]{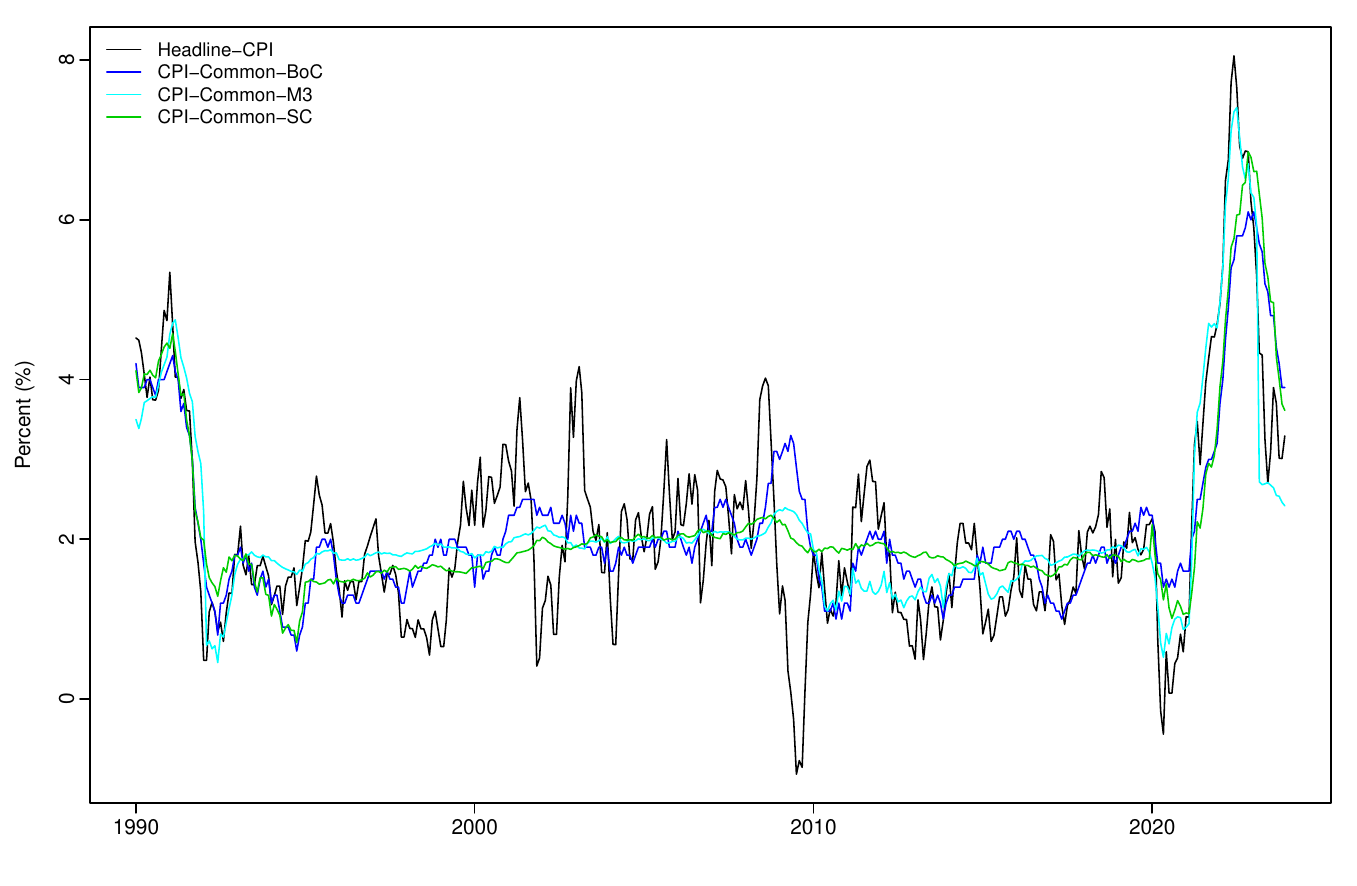}
    \label{fig:cpicommon_sc_ms}
    \parbox[t]{0.95\linewidth}{%
    \scriptsize
    \raggedright 
    \textbf{Notes}: The time series include Headline CPI (black), CPI-Common (blue), the Bank of Canada’s time-invariant factor model, CPI-Common-M3 (cyan), using the Markov switching approach proposed here with $M=3$ regimes, and CPI-Common-SC (green), using the structural break approach proposed here.}
\end{figure}
Finally, Figure \ref{fig:cpicommon_sc_ms} shows headline inflation, the BoC's CPI-Common, and our two proposed measures. Specifically, the CPI-Common-SC, which incorporates the structural breaks approach, and the CPI-Common-MS, which uses our Markov switching approach with $M=3$ regimes. We present all three indicators for the full sample, from January 1990 to December 2023. Both of our proposed measures are more stable throughout the sample, as their time-varying nature allows them to better capture the stability of inflation during periods when inflation is close to the target. Furthermore, both indicators track headline inflation more closely when it begins to deviate from the target. Compared to the CPI-Common-SC, the CPI-Common-MS once again presents a more optimistic view at the end of our sample, suggesting that inflation is now much closer to the Bank's target.

%%%%%%%%%%%%%%%%%%%%%%%%%%%%%%%%%%%%%%%%%%%%%%%%%%%%%%%%%%%%%%%%%%%%%%%%%%%%%%%%%
%%.                               CONCLUSION                                   %%
%%%%%%%%%%%%%%%%%%%%%%%%%%%%%%%%%%%%%%%%%%%%%%%%%%%%%%%%%%%%%%%%%%%%%%%%%%%%%%%%%
% --------------------------------------
\section{Conclusion}
% --------------------------------------
This paper proposes two novel indicators for underlying core inflation that are robust to high frequency volatility from transitory and sector-specific shocks and to abrupt changes in inflation.  This robustness to abrupt changes is a unique aspect of the indicators we propose, developed within a time-varying framework. The first indicator adopts a structural change approach, which is highly flexible but limited by its off-line nature. The second indicator is constructed using a Markov-switching framework, providing a real-time indicator of underlying core inflation that we argue can be useful as a short-term guide for monetary policy. 

To demonstrate the effectiveness of our proposed indicators, we apply them to Canadian price data, proposing the CPI-Common-SC (based on structural change) and the CPI-Common-MS (using Markov switching) as potential substitutes for the BoC’s CPI-Common. Both methods succeed in eliminating or reducing revisions, an issue that has intensified amid recent inflation surge. These large CPI-Common revisions have cast doubt on the reliability of the current indicator, and our methods seek to address these challenges. Additionally, we evaluate the real-time performance of the CPI-Common-MS against a benchmark model without Markov switching, finding that the Markov-switching approach outperforms in terms of predictive accuracy for headline inflation over the short and medium term. Importantly, estimating the CPI-Common-MS also provides the probability of being in each regime at each point in time. 

Despite these advancements, there are important directions for future research. For the Markov-switching model, developing a formal hypothesis testing procedure is essential to determine the appropriate number of regimes at any given time, $t$. This could potentially lead to further improvements in the real-time performance of this indicator. Furthermore, additional research is needed to evaluate the stability of the number of factors across different regimes. Such investigations are currently in progress.
    
%%%%%%%%%%%%%%%%%%%%%%%%%%%%%%%%%%%%%%%%%%%%%%%%%%%%%%%%%%%%%%%%%%%%%%%%%%%%%%%%%
%%                                REFERENCES                                   %%
%%%%%%%%%%%%%%%%%%%%%%%%%%%%%%%%%%%%%%%%%%%%%%%%%%%%%%%%%%%%%%%%%%%%%%%%%%%%%%%%%
%\newpage
% --------------------------------------
\addcontentsline{toc}{section}{References}
% --------------------------------------

%\printbibliography
{
  \linespread{1.00}\selectfont
  \bibliographystyle{apalike}
  \bibliography{sc_core_inf_lib.bib}
}
\linespread{1.00}\selectfont

%%%%%%%%%%%%%%%%%%%%%%%%%%%%%%%%%%%%%%%%%%%%%%%%%%%%%%%%%%%%%%%%%%%%%%%%%%%%%%%%%
%%                                 APPENDIX                                    %%
%%%%%%%%%%%%%%%%%%%%%%%%%%%%%%%%%%%%%%%%%%%%%%%%%%%%%%%%%%%%%%%%%%%%%%%%%%%%%%%%%
% --------------------------------------
\addcontentsline{toc}{section}{Appendix}
% --------------------------------------
\appendix
\counterwithin{figure}{section}
\counterwithin{table}{section}

\newpage
\section{Additional Tables}
\begin{table}[H]
\setlength\tabcolsep{6.5pt}
\caption{Root Mean Absolute Difference of real-time vs. full-information for each model}
\label{tab:real_vs_fullinfo_mad}
\begin{center}
\begin{scriptsize}
\begin{tabular}{lccccccccccccccc}
\hline\hline
 & \multicolumn{2}{c}{Pre-COVID} & & \multicolumn{2}{c}{Rising Inflation} & & \multicolumn{2}{c}{Inflation Norm.} & & \multicolumn{2}{c}{Post-COVID} & &\multicolumn{2}{c}{Full-Sample} \bigstrut \\ 
\hhline{~ - - ~ - - ~ - - ~ - - ~ - -}
                   & Y-o-Y & M-o-M & & Y-o-Y & M-o-M & & Y-o-Y & M-o-M & & Y-o-Y & M-o-M & & Y-o-Y & M-o-M \bigstrut \\
\hline
$\hat{\pi}^{M1}_t$ & 0.073 & 0.073 & & 0.888 & 0.056 & & 0.119 & 0.034 & & 0.486 & 0.060 & & 0.267 & 0.070 \\
$\hat{\pi}^{M2}_t$ & 0.220 & 0.058 & & 1.051 & 0.092 & & 0.069 & 0.013 & & 0.541 & 0.073 & & 0.323 & 0.090 \\
$\hat{\pi}^{M3}_t$ & 0.294 & 0.082 & & 0.632 & 0.119 & & 0.083 & 0.031 & & 0.225 & 0.106 & & 0.270 & 0.084 \\
$\hat{\pi}^{M4}_t$ & 0.310 & 0.084 & & 0.607 & 0.078 & & 0.469 & 0.088 & & 0.786 & 0.118 & & 0.391 & 0.081 \\
\hline\hline
\end{tabular}%     
\end{scriptsize}
\parbox[t]{0.95\linewidth}{%
\scriptsize
\raggedright 
\textbf{Notes}: Y-o-Y is Year-over-Year while M-o-M is Month-over-Month inflation rates. In this table, we compare real-time estimates against the full information estimates for each model. That is, we use the difference $\hat{\pi}_t^{m,r} - \hat{\pi}_t^{m,f}$ for each model $m=\{M1, M2, M3, M4\}$ and for each sample. Values shown are root mean absolute differences. The Pre-COVID sample ranges from 1990-01 to 2019-12, Rising Inflation sample ranges from 2020-01 to 2022-12, Inflation Normalization sample ranges from 2023-01 to 2023-12, Post-COVID sample ranges from 2020-01 to 2023-12, and Full-Sample ranges from 1990-01 to 2023-12.}
\end{center}
\end{table}

\begin{table}[H]
\setlength\tabcolsep{7pt}
\caption{Forecasting Headline CPI $\pi^{\text{HCPI}}_{t+h}$ using $T_{R}$ statistic}
\label{tab:forecasting_tr_M75}
\begin{center}
\begin{scriptsize}
\begin{tabular}{lccccccccccc}
\hline\hline
\multirow{2}{*}{Models} & \multicolumn{5}{c}{Year-over-Year}  & & \multicolumn{5}{c}{Month-over-Month} \bigstrut\\
\cline{2-6} \cline{8-12}
 & $h=1$ &  $h=3$ &  $h=6$ &  $h=9$ &  $h=12$ &   & $h=1$ &  $h=3$ &  $h=6$ &  $h=9$ &  $h=12$  \bigstrut \\  
\hline
& \multicolumn{11}{c}{$h$-step ahead from 2020-01 to 2023-12 ($T_f=48$)} \bigstrut\\
\hline
\multirow{2}{*}{$\hat{\pi}^{M1}_t$} &                           1.823  &                           2.157  &                           2.624  &                  \textbf{2.914}  &                  \textbf{3.093}   &   &                  \textbf{0.601}   &                           0.653   &                  \textbf{0.676}   &                  \textbf{0.596}   &                           0.520  \\
                                    &                         (0.019)  &                         (0.016)  &                         (0.122)  &                         (0.366)  &                         (0.572)   &   &                         (0.936)   &                         (0.000)   &                         (0.769)   &                         (0.671)   &                         (0.236)  \\
\multirow{2}{*}{$\hat{\pi}^{M2}_t$} &                           1.626  &                           2.041  &                           2.576  &                  \textbf{2.973}  &                           3.232   &   & \textbf{\textcolor{red}{0.591}}   &                  \textbf{0.597}   &                  \textbf{0.651}   &                  \textbf{0.580}   &                  \textbf{0.492}  \\
                                    &                         (0.029)  &                         (0.031)  &                         (0.134)  &                         (0.253)  &                         (0.131)   &   &                         (1.000)   &                         (0.891)   &                         (0.985)   &                         (0.709)   &                         (0.997)  \\
\multirow{2}{*}{$\hat{\pi}^{M3}_t$} &                  \textbf{1.163}  &                  \textbf{1.542}  &                  \textbf{2.108}  &                  \textbf{2.592}  &                  \textbf{2.978}   &   &                  \textbf{0.594}   & \textbf{\textcolor{red}{0.594}}   &                  \textbf{0.653}   &                  \textbf{0.594}   &                  \textbf{0.493}  \\
                                    &                         (0.285)  &                         (0.272)  &                         (0.294)  &                         (0.620)  &                         (0.330)   &   &                         (0.981)   &                         (1.000)   &                         (0.979)   &                         (0.637)   &                         (0.988)  \\
\multirow{2}{*}{$\hat{\pi}^{M4}_t$} & \textbf{\textcolor{red}{1.130}}  & \textbf{\textcolor{red}{1.509}}  & \textbf{\textcolor{red}{2.073}}  & \textbf{\textcolor{red}{2.545}}  & \textbf{\textcolor{red}{2.923}}   &   &                  \textbf{0.596}   &                  \textbf{0.610}   & \textbf{\textcolor{red}{0.646}}   & \textbf{\textcolor{red}{0.555}}   & \textbf{\textcolor{red}{0.491}}  \\
                                    &                         (1.000)  &                         (1.000)  &                         (1.000)  &                         (1.000)  &                         (1.000)   &   &                         (0.982)   &                         (0.645)   &                         (1.000)   &                         (1.000)   &                         (1.000)  \\
\hline
& \multicolumn{11}{c}{$h$-step ahead from 2007-01 to 2023-12 ($T_f=204$)} \bigstrut\\
\hline
\multirow{2}{*}{$\hat{\pi}^{M1}_t$} &                           1.176  &                           1.328  &                           1.537  &                  \textbf{1.667}  &                  \textbf{1.742}   &   &                           0.471   &                           0.504   &                  \textbf{0.502}   &                  \textbf{0.479}   &                  \textbf{0.386}  \\
                                    &                         (0.158)  &                         (0.189)  &                         (0.215)  &                         (0.502)  &                         (0.847)   &   &                         (0.080)   &                         (0.022)   &                         (0.268)   &                         (0.100)   &                         (0.916)  \\
\multirow{2}{*}{$\hat{\pi}^{M2}_t$} &                  \textbf{1.053}  &                  \textbf{1.245}  &                  \textbf{1.498}  &                  \textbf{1.685}  &                  \textbf{1.796}   &   & \textbf{\textcolor{red}{0.452}}   &                  \textbf{0.489}   & \textbf{\textcolor{red}{0.481}}   &                           0.475   & \textbf{\textcolor{red}{0.379}}  \\
                                    &                         (0.082)  &                         (0.096)  &                         (0.123)  &                         (0.449)  &                         (0.476)   &   &                         (1.000)   &                         (0.367)   &                         (1.000)   &                         (0.043)   &                         (1.000)  \\
\multirow{2}{*}{$\hat{\pi}^{M3}_t$} &                           0.892  &                           1.062  &                           1.320  &                  \textbf{1.535}  &                  \textbf{1.697}   &   &                  \textbf{0.456}   &                           0.496   &                  \textbf{0.490}   &                           0.484   &                  \textbf{0.381}  \\
                                    &                         (0.174)  &                         (0.152)  &                         (0.171)  &                         (0.303)  &                         (0.716)   &   &                         (0.780)   &                         (0.150)   &                         (0.682)   &                         (0.072)   &                         (0.999)  \\
\multirow{2}{*}{$\hat{\pi}^{M4}_t$} & \textbf{\textcolor{red}{0.833}}  & \textbf{\textcolor{red}{1.012}}  & \textbf{\textcolor{red}{1.280}}  & \textbf{\textcolor{red}{1.502}}  & \textbf{\textcolor{red}{1.673}}   &   &                  \textbf{0.459}   & \textbf{\textcolor{red}{0.480}}   &                  \textbf{0.485}   & \textbf{\textcolor{red}{0.458}}   &                  \textbf{0.383}  \\
                                    &                         (1.000)  &                         (1.000)  &                         (1.000)  &                         (1.000)  &                         (1.000)   &   &                         (0.539)   &                         (1.000)   &                         (0.953)   &                         (1.000)   &                         (0.963)  \\
\hline\hline
\end{tabular}%     
\end{scriptsize}
\parbox[t]{0.95\linewidth}{%
\scriptsize
\raggedright 
\textbf{Notes}: Reported values are the root mean squared forecast error of model $m$ at horizon $h$ (i.e., $\text{RMSFE}^m_h$). Estimation of $\hat{\pi}^{m}_t$ is performed using an expanding window with information at time $t$ only (real-time). Using the $T_{R}$ test of \cite{Hansen-Lunde-Nason2011} with $\alpha=0.25$. Values in bold highlight models that belong to $\hat{\mathcal{M}}_{75\%}$ and values in \textcolor{red}{red} highlight those that belong to the MCS and have lowest RMSFE. The values in parenthesis are the associated p-values of the MCS test using the sequential algorithm. The p-values of the eliminated models are those from the previous iteration when model was eliminated.}
\end{center}
\end{table}

\section{Additional Figures}
%\vspace{-2em}
\begin{figure}[H]
    \centering
    \caption{Difference in April vs. December 2022 estimates with and without SC}
    \includegraphics[width=0.90\textwidth]{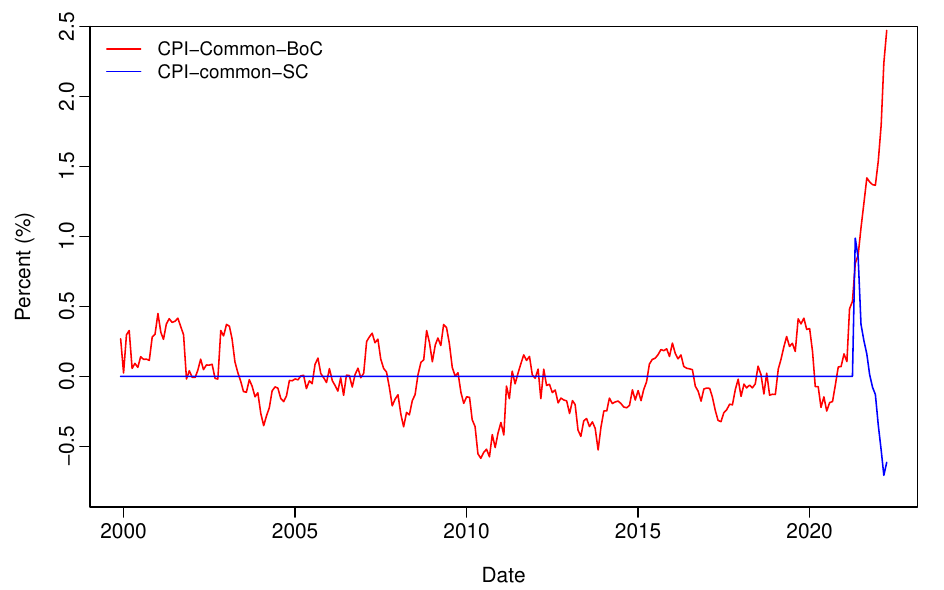}
    \label{fig:worstrevdiff_with_sc}
    \parbox[t]{0.95\linewidth}{%
    \scriptsize
    \raggedright 
    \textbf{Notes}: This chart plots the difference between the blue and red line of Figure \ref{fig:worstrev_with_without_sc} (time-invariant benchmark revisions) in red and the difference between the black and green line of Figure \ref{fig:worstrev_with_without_sc} (structural break approach revisions) in blue.}
\end{figure}

%\vspace{-2.5em}
\begin{figure}[H]
    \centering
    \caption{The $h$-step ahead RMSFE from January 2020 to December 2023 ($T_f=48$)}
    \includegraphics[width=0.90\textwidth]{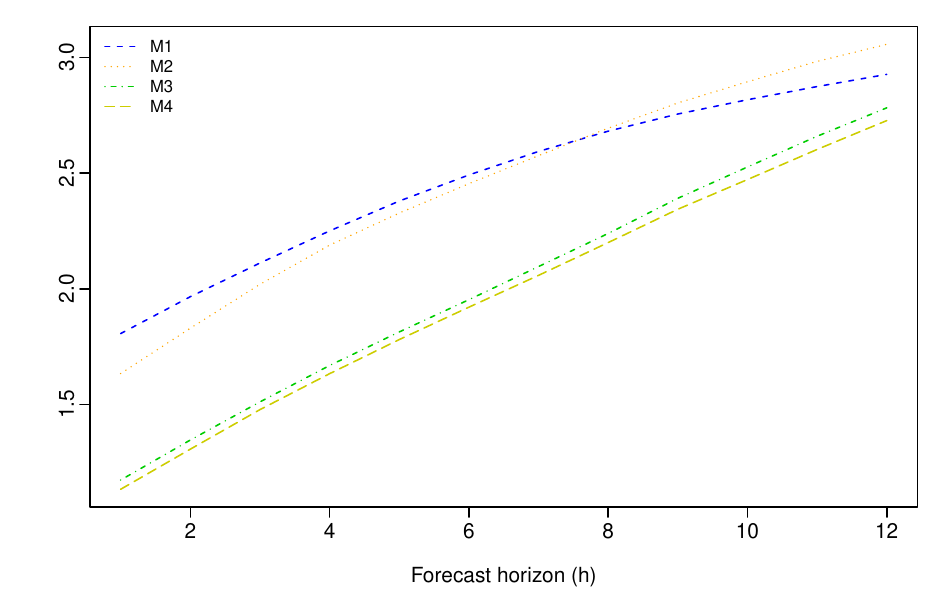}
    \label{fig:realtime_fcast_2020}
    \parbox[t]{0.95\linewidth}{%
    \scriptsize
    \raggedright 
    \textbf{Notes}: RMSFE from horizons $h=1$ to horizon $h=12$ in steps of $1$ month for CPI-Common (benchmark) in blue, CPI-Common-M2 (Markov switching with $M=2$) in orange, CPI-Common-M3 (Markov switching with $M=3$) in green, and CPI-Common-M4 (Markov switching with $M=4$) in yellow. These values are obtained by using the forecasting window from January 2020 to December 2023 such that $T_{f}=48$.}
\end{figure}

\begin{figure}[H]
    \centering
    \caption{The $h$-step ahead RMSFE from January 2007 to December 2023 ($T_f=204$)}
    \includegraphics[width=0.90\textwidth]{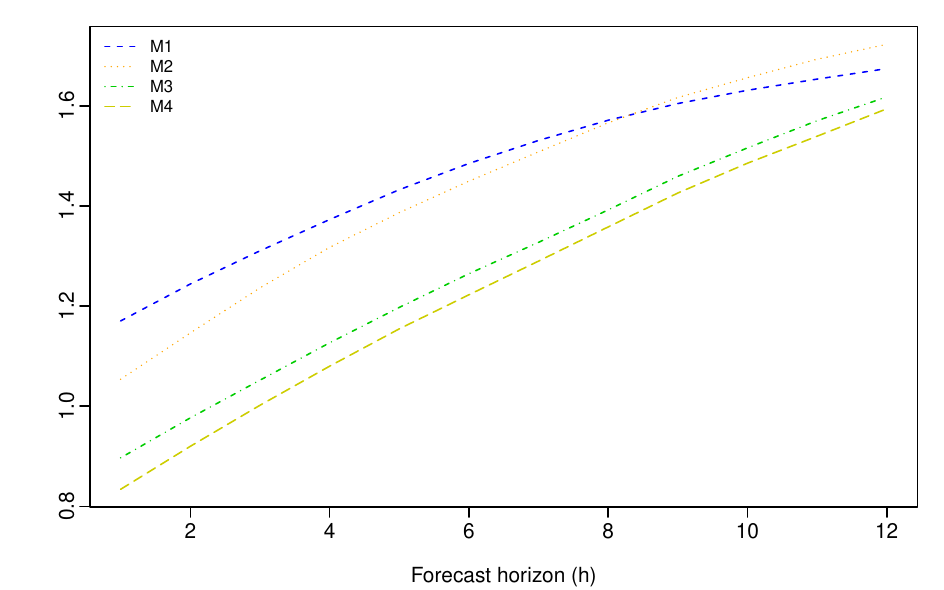}
    \label{fig:realtime_fcast_2007}
    \parbox[t]{0.95\linewidth}{%
    \scriptsize
    \raggedright 
    \textbf{Notes}: RMSFE from horizons $h=1$ to horizon $h=12$ in steps of $1$ month for CPI-Common (benchmark) in blue, CPI-Common-M2 (Markov switching with $M=2$) in orange, CPI-Common-M3 (Markov switching with $M=3$) in green, and CPI-Common-M4 (Markov switching with $M=4$) in yellow. These values are obtained by using the forecasting window from January 2007 to December 2023 such that $T_{f}=204$.}
\end{figure}

\begin{figure}[H]
    \centering
    \caption{All estimates in real-time from January 1990 to December 2023}
    \includegraphics[width=0.90\textwidth]{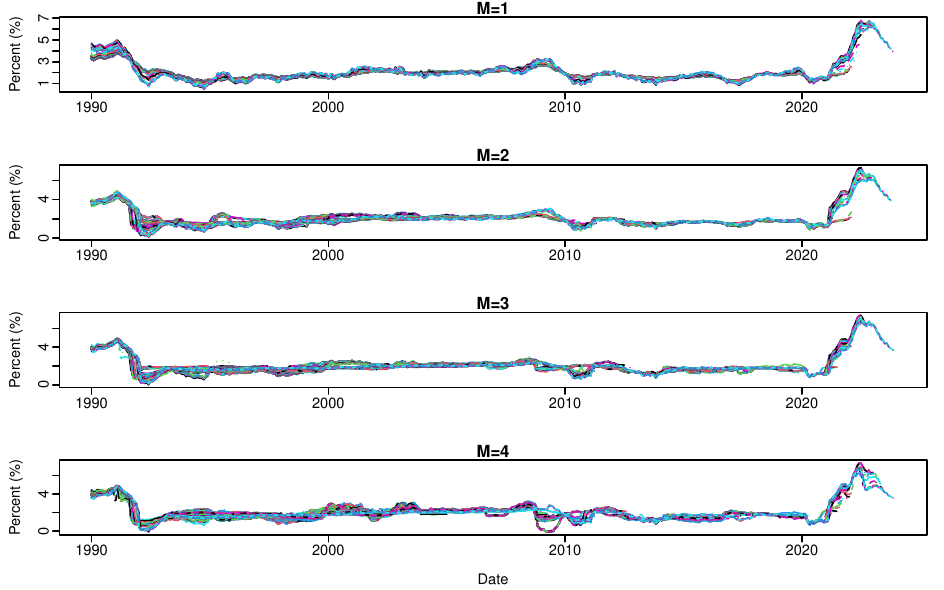}
    \label{fig:realtime_rev_1990}
    \parbox[t]{0.95\linewidth}{%
    \scriptsize
    \raggedright 
    \textbf{Notes}: Top Chart: Real-time stacked estimates of CPI-Common (benchmark). First Middle Chart: Real-time stacked estimates of CPI-Common-M2 (Markov switching with $M=2$). Second Middle Chart: Real-time stacked estimates of CPI-Common-M3 ($M=3$). Bottom Chart: Real-time stacked estimates of CPI-Common-M4 ($M=4$). Greater spread across estimates for a given date $t$ indicates larger revisions for that date.}
\end{figure}

\begin{figure}[H]
    \centering
    \caption{All estimates in real-time from January 2020 to December 2023}
    \includegraphics[width=0.90\textwidth]{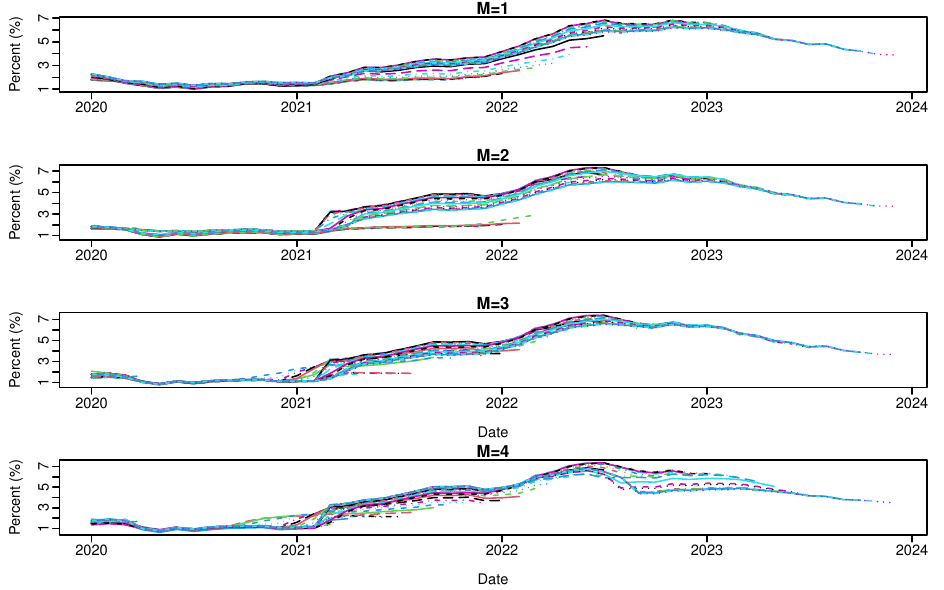}
    \label{fig:realtime_rev_2020}
    \parbox[t]{0.95\linewidth}{%
    \scriptsize
    \raggedright 
    \textbf{Notes}: Top Chart: Real-time stacked estimates of CPI-Common (benchmark). First Middle Chart: Real-time stacked estimates of CPI-Common-M2 (Markov switching with $M=2$). Second Middle Chart: Real-time stacked estimates of CPI-Common-M3 ($M=3$). Bottom Chart: Real-time stacked estimates of CPI-Common-M4 ($M=4$). Greater spread across estimates for a given date $t$ indicates larger revisions for that date. This chart only includes data from January 2020 to December 2023 for clarity.}
\end{figure}

\begin{comment}
\begin{figure}[!tbh]
    \centering
    \includegraphics[width=0.8\textwidth]{figures/cpi_common_worstrev_diff_boc_ms_20221201.pdf}
    \caption{\footnotesize{Difference in estimates (December 2022 - April 2022)}}
    \label{fig:worstrevdiff_ms}
\end{figure}

\begin{figure}[!tbh]
    \centering
    \includegraphics[width=0.8\textwidth]{figures/cpi_common_cad_dist_HeadInf_h_YoY_20200101_20221201.pdf}
    \caption{\footnotesize{}}
    \label{fig:realtime_fcast}
\end{figure}

\begin{figure}[!tbh]
    \centering
    \includegraphics[width=0.8\textwidth]{figures/cpi_common_cad_dist_HeadInf_h_YoY_20221201.pdf}
    \caption{\footnotesize{}}
    \label{fig:realtime_fcast}
\end{figure}

\begin{figure}[!tbh]
    \centering
    \includegraphics[width=0.8\textwidth]{figures/cpi_common_cad_dist_HeadInf_h_MoM_20200101_20221201.pdf}
    \caption{\footnotesize{}}
    \label{fig:realtime_fcast}
\end{figure}

\begin{figure}[!tbh]
    \centering
    \includegraphics[width=0.8\textwidth]{figures/cpi_common_cad_dist_HeadInf_h_MoM_20221201.pdf}
    \caption{\footnotesize{}}
    \label{fig:realtime_fcast}
\end{figure}

\begin{figure}[!tbh]
    \centering
    \includegraphics[width=0.8\textwidth]{figures/cpi_common_cad_dist_HeadInf_h_YoY_20200101_20230601_poster.pdf}
    \caption{\footnotesize{}}
    \label{fig:realtime_fcast}
\end{figure}

\end{comment}

\end{document}